\documentclass[zpreprint,zbstepj]{zeus_paper}

\usepackage[english]{babel}

\chardef\usc=95
\chardef\til=126
\catcode`\@=11 
\DeclareRobustCommand\xdotspace{\futurelet\@let@token\@xdotspace}
\def\@xdotspace{%
  \ifx\@let@token.\else
  \ifx\@let@token\bgroup.\else
  \ifx\@let@token\egroup.\else
  \ifx\@let@token\/.\else
  \ifx\@let@token\ .\else
  \ifx\@let@token~.\else
  \ifx\@let@token!.\else
  \ifx\@let@token,.\else
  \ifx\@let@token:.\else
  \ifx\@let@token;.\else
  \ifx\@let@token?.\else
  \ifx\@let@token/.\else
  \ifx\@let@token'.\else
  \ifx\@let@token).\else
  \ifx\@let@token-.\else
  \ifx\@let@token\@xobeysp.\else
  \ifx\@let@token\space.\else
  \ifx\@let@token\@sptoken.\else
   .\space
   \fi\fi\fi\fi\fi\fi\fi\fi\fi\fi\fi\fi\fi\fi\fi\fi\fi\fi}
\catcode`\@=12 

\newcommand{\stru}[2]{%
   \relax\ifmmode\hbox{\vrule height#1 depth#2 width0pt}%
   \else\vrule height#1 depth#2 width0pt\fi}

\newcommand{\Ronum}[1]{\uppercase\expandafter{\romannumeral#1}}
\newcommand{\ronum}[1]{\expandafter{\romannumeral#1}}
\DeclareRobustCommand{\LaTeXZ}{%
  \LaTeX\kern-.05em4\kern-.1em
  {\raisebox{-0.2ex}{$\scriptstyle\text{ZEUS}$}}\xspace}

\DeclareMathAlphabet{\mathbf}{OT1}{cmr}{bx}{sl}
\newcommand{\eVdist}{\kern-0.06667em}

\newcommand{\Gev}{{\text{Ge}\eVdist\text{V\/}}}

\newcommand{\gev}{{\,\text{Ge}\eVdist\text{V\/}}}

\newcommand{\pb}{\,\text{pb}}
\newcommand{\fb}{\,\text{fb}}

\newcommand{\Tesla}{\,\text{T}}

\newcommand{\slashfrac}[2]{%
  \raisebox{0.5ex}{\ensuremath #1}\kern-0.12em/\kern-0.08em
  \raisebox{-.8ex}{\ensuremath #2}}

\newcommand{\sqr}[3]{%
    {\vcenter{\hrule height.#3ex\hbox{\vrule width.#2ex height#1ex
     \kern#1ex\vrule width.#3ex}\hrule height.#2ex}}}

\catcode`\@=11 
\newcommand{\parenbar}{\mathpalette\p@renb@r}
\def\p@renb@r#1#2{\vbox{%
  \ifx#1\scriptscriptstyle \dimen@.7em\dimen@ii.2em\else
  \ifx#1\scriptstyle \dimen@.8em\dimen@ii.25em\else
  \dimen@1em\dimen@ii.4em\fi\fi \offinterlineskip
  \ialign{\hfill##\hfill\cr
    \vbox{\hrule width\dimen@ii}\cr
    \noalign{\vskip-.3ex}%
    \hbox to\dimen@{$\mathchar300\hfil\mathchar301$}\cr
    \noalign{\vskip-.3ex}%
    $#1#2$\cr}}}
\catcode`\@=12 

\newcommand{\IP}{{\rm I$\kern-0.01667em$P}\xspace}

\newcommand{\F}{{\cal F}}

\mathchardef\qsm=63
\mathchardef\pls=43
\mathchardef\mns=512
\mathchardef\plm=518
\mathchardef\eql=61
\mathchardef\smallleft=300
\mathchardef\smallright=301
\mathchardef\les=316
\mathchardef\gre=318
\mathchardef\leq=532
\mathchardef\grq=533

\catcode`\@=11 
\newcounter{pict@width}
\newcounter{pict@height}
\newlength{\pict@scale}
\newcommand{\psfigadd}[4]{%
\setcounter{pict@width}{1*\ratio{#2+\pict@scale/2}{\pict@scale}}
\setcounter{pict@height}{1*\ratio{#3+\pict@scale/2}{\pict@scale}}
\setlength{\unitlength}{\pict@scale}
\hbox to #2{\hspace{-\fill}\begin{picture}(\thepict@width,\thepict@height)
\put(0,0){\psfig{figure=#1,width=#2,height=#3,clip=}}
\SetScale{0.283466457}
\SetWidth{1.763889}
{#4}
\end{picture}}
}
\newcounter{pict@widthfst}
\newcounter{pict@widthscd}
\newcounter{pict@widthtot}
\newcommand{\psfigaddtwo}[7]{%
\setcounter{pict@widthfst}{1*\ratio{#2+\pict@scale/2}{\pict@scale}}
\setcounter{pict@widthscd}{1*\ratio{#2+#4+\pict@scale/2}{\pict@scale}}
\setcounter{pict@widthtot}{1*\ratio{#2+#4+#6+\pict@scale/2}{\pict@scale}}
\setcounter{pict@height}{1*\ratio{#3+\pict@scale/2}{\pict@scale}}
\setlength{\unitlength}{\pict@scale}
\hbox{\hspace{-\fill}\begin{picture}(\thepict@widthtot,\thepict@height)
\put(0,0){\psfig{figure=#1,width=#2,height=#3,clip=}}
\put(\thepict@widthscd,0){\psfig{figure=#5,width=#6,height=#3,clip=}}
\SetScale{0.283466457}
\SetWidth{1.763889}
{#7}
\end{picture}}
}
\newcommand{\psfigror}[4]{%
\setcounter{pict@width}{1*\ratio{#2+\pict@scale/2}{\pict@scale}}
\setcounter{pict@height}{1*\ratio{#3+\pict@scale/2}{\pict@scale}}
\setlength{\unitlength}{\pict@scale}
\hbox{\begin{picture}(\thepict@width,\thepict@height)
\put(0,\thepict@height){\psfig{figure=#1,width=#3,height=#2,clip=,angle=270}}
\SetScale{0.283466457}
\SetWidth{1.763889}
{#4}
\end{picture}}
}
\newcommand{\psfigrol}[4]{%
\setcounter{pict@width}{1*\ratio{#2+\pict@scale/2}{\pict@scale}}
\setcounter{pict@height}{1*\ratio{#3+\pict@scale/2}{\pict@scale}}
\setlength{\unitlength}{\pict@scale}
\hbox{\begin{picture}(\thepict@width,\thepict@height)
\put(0,0){\psfig{figure=#1,width=#3,height=#2,clip=,angle=90}}
\SetScale{0.283466457}
\SetWidth{1.763889}
{#4}
\end{picture}}
}
\catcode`\@=12 
\newlength\listtextwidth

\catcode`\@=11 
\newlength{\@tabfninsert}
\newlength{\@tabfnwidth}
\newcommand{\tabfootnote}[2]{%
  \setlength{\@tabfninsert}{0.8em}
  \setlength{\@tabfnwidth}{\textwidth}
  \addtolength{\@tabfnwidth}{-\@tabfninsert}
  \addtolength{\@tabfnwidth}{-0.4em}
  \noindent\makebox[\@tabfninsert][r]{\footnotesize$^{#1}$\hfil}\hfill%
  \parbox[t]{\@tabfnwidth}{\footnotesize #2\hfill}}
\catcode`\@=12 

\def\JHEP{JHEP}

\def\etjet{E_T^{\rm jet}}
\def\etajet{\eta^{\rm jet}}
\def\phijet{\phi^{\rm jet}}
\def\etsbj{E_T^{\rm sbj}}
\def\etasbj{\eta^{\rm sbj}}
\def\phisbj{\phi^{\rm sbj}}
\def\asbj{\alpha^{\rm sbj}}

\def\etar{-1<\etajet<2.5}

\def\qg2{$\q2>125$~\g2}

\def\sq2{d\sigma/d\q2}
 
\def\q2{Q^2}

\def\pb1{pb$^{-1}$}
\def\fb1{fb$^{-1}$}

\def\g2{GeV$^2$}

\def\F2g{F_2^{\gamma}}
\def\f2gv{F_2^{\gamma^*}}

\def\rr1{R=1}
\def\r7{R=0.7}
\def\R71{R=0.7\ {\rm and}\ 1}

\def\m3j{M^{\rm 3j}}

\def\kt{k_T}

\def\lq2{\log_{10}(\q2)}

\def\qq{q\bar q}

\def\colab#1{#1 Coll.}

\def\ptmis{p_T^{\rm miss}}

\def\yc{y_{\rm cut}}

\def\z0{Z^0}
\def\mz{M_Z}

\def\as{\alpha_s}
\def\oalphas2{{\cal O}(\alpha\as^2)}

\def\oass{{\cal O}(\as^2)}

\def\asz{\as(\mz)}

\def\p2{P^2}

\def\mr2{\mu_R^2}
\def\mf2{\mu_F^2}

\def\etal{et al.}

\def\bet0#1#2#3#4#5#6{\beta_0 = #1\pm #2\ {\rm (stat.)}\ ^{+#4}_{-#3}\ {\rm (exp.)}\ ^{+#6}_{-#5}\ {\rm (th.)}}

\def\a34{\alpha_{23}}

\def\figdir{./}

\begin{document}
\prepnum{{DESY--08--178}}

\title{
            Subjet distributions in
              deep inelastic scattering at HERA
}                                                       
                    
\author{ZEUS Collaboration}
\date{December 2008}

\abstract{
Subjet distributions were measured in neutral current deep
inelastic $ep$ scattering with the ZEUS detector at HERA using an
integrated luminosity of 81.7~\pb1. Jets were identified using the
$\kt$ cluster algorithm in the laboratory frame. Subjets were defined
as jet-like substructures identified by a reapplication of the cluster
algorithm at a smaller value of the resolution parameter
$\yc$. Measurements of subjet distributions for jets with exactly two
subjets for $\yc=0.05$ are presented as functions of observables
sensitive to the pattern of parton radiation and to the colour
coherence between the initial and final states. Perturbative QCD
predictions give an adequate description of the data.
}

\makezeustitle

\def\3{\ss}

\pagenumbering{Roman}

\begin{center}

{                      \Large  The ZEUS Collaboration              }

\end{center}

  S.~Chekanov,
  M.~Derrick,
  S.~Magill,
  B.~Musgrave,
  D.~Nicholass$^{   1}$,
  \mbox{J.~Repond},
  R.~Yoshida\\
 {\it Argonne National Laboratory, Argonne, Illinois 60439-4815,
   USA}~$^{n}$
\par \filbreak

  M.C.K.~Mattingly \\
 {\it Andrews University, Berrien Springs, Michigan 49104-0380, USA}
\par \filbreak

  P.~Antonioli,
  G.~Bari,
  L.~Bellagamba,
  D.~Boscherini,
  A.~Bruni,
  G.~Bruni,
  F.~Cindolo,
  M.~Corradi,
\mbox{G.~Iacobucci},
  A.~Margotti,
  R.~Nania,
  A.~Polini\\
  {\it INFN Bologna, Bologna, Italy}~$^{e}$
% ...

\newpage

\enlargethispage{5cm}

$^{\    1}$ also affiliated with University College London,
United Kingdom\\
% ...
$^{\    5}$ supported by the research grant no. 1 P03B 04529
(2005-2008) \\
% ...

\newpage

\begin{tabular}[h]{rp{14cm}}

$^{a}$ &  supported by the Natural Sciences and Engineering Research
Council of Canada (NSERC) \\
$^{b}$ &  supported by the German Federal Ministry for Education and
Research (BMBF), under contract numbers 05 HZ6PDA, 05 HZ6GUA, 05
HZ6VFA and 05 HZ4KHA\\
$^{c}$ &  supported in part by the MINERVA Gesellschaft f\"ur
Forschung GmbH, the Israel Science Foundation (grant no. 293/02-11.2)
and the U.S.-Israel Binational Science Foundation \\
$^{d}$ &  supported by the Israel Science Foundation\\
$^{e}$ &  supported by the Italian National Institute for Nuclear
Physics (INFN) \\
$^{f}$ &  supported by the Japanese Ministry of Education, Culture,
Sports, Science and Technology (MEXT) and its grants for Scientific
Research\\
$^{g}$ &  supported by the Korean Ministry of Education and Korea
Science and Engineering Foundation\\
$^{h}$ &  supported by the Netherlands Foundation for Research on
Matter (FOM)\\
$^{i}$ &  supported by the Polish State Committee for Scientific
Research, project no. DESY/256/2006 - 154/DES/2006/03\\
$^{j}$ &  partially supported by the German Federal Ministry for
Education and Research (BMBF)\\
$^{k}$ &  supported by RF Presidential grant N 1456.2008.2 for the
leading scientific schools and by the Russian Ministry of Education
and Science through its grant for Scientific Research on High Energy
Physics\\
$^{l}$ &  supported by the Spanish Ministry of Education and Science
through funds provided by CICYT\\
$^{m}$ &  supported by the Science and Technology Facilities Council,
UK\\
$^{n}$ &  supported by the US Department of Energy\\
$^{o}$ &  supported by the US National Science Foundation. Any
opinion, findings and conclusions or recommendations expressed in this
material are those of the authors and do not necessarily reflect the
views of the National Science Foundation.\\
$^{p}$ &  supported by the Polish Ministry of Science and Higher
Education as a scientific project (2006-2008)\\
$^{q}$ &  supported by FNRS and its associated funds (IISN and FRIA)
and by an Inter-University Attraction Poles Programme subsidised by
the Belgian Federal Science Policy Office\\
$^{r}$ &  supported by an FRGS grant from the Malaysian government\\

\end{tabular}

\newpage

\pagenumbering{arabic} 
\pagestyle{plain}

\section{Introduction}

Jet production in $ep$ collisions provides a wide testing ground
of perturbative QCD (pQCD). Measurements of differential cross
sections for jet 
production~\cite{pl:b531:9,*epj:c23:615,pl:b443:394,*pl:b507:70,*pl:b547:164,*epj:c23:13,*pl:b560:7,*epj:c31:149,*np:b765:1,*pl:b649:12,pl:b515:17,*epj:c19:289,*epj:c19:429,*epj:c25:13,*pl:b542:193,*epj:c29:497,*pl:b639:21,*pl:b653:134}
have allowed detailed studies of parton dynamics, tests of the proton
and photon parton distribution functions (PDFs) as well as precise
determinations of the strong coupling constant, $\as$. 

Gluon emission from primary quarks was
investigated~\cite{pl:b558:41,np:b700:3} by means of the
internal structure of jets; these type of studies gave insight into
the transition between a parton produced in a hard process and the
experimentally observable jet of hadrons. The pattern of parton
radiation within a jet is dictated in QCD by the splitting
functions. These functions, $P_{ab}(z,\mu)$ with $a,b=q$ or $g$, are
interpreted as the probability that a parton of type $b$, having
radiated a parton of type $a$, is left with a fraction $z$ of the
longitudinal momentum of the parent parton and a transverse momentum
squared smaller than $\mu^2$, where $\mu$ is the typical hard scale of
the process. The splitting functions are calculable as power series in
$\as$. Thus, the characteristics of jet substructure provide direct
access to the QCD splitting functions and their dependence on the scale.

The understanding of jet substructure is also important in the context
of jet identification in boosted systems, like hadronic top
decays~\cite{jhep:0807:092,*hep-ph:0806.0848} or $b\bar b$ final
states at LHC~\cite{prl:100:242001}. The first example calls for a
direct application of jet substructure, the second requires knowledge
about jet substructure to distinguish between single- and double-quark
induced jets. This paper presents a study of jet substructure in a
more controlled hadronic-type environment than that provided by
hadron-hadron colliders.

Jet production in neutral current (NC) deep inelastic scattering (DIS)
was previously used to study the mean subjet
multiplicity~\cite{pl:b558:41} and the mean integrated jet
shape~\cite{np:b700:3} with values of $\asz$ extracted from those
measurements. In the present study, the pattern of QCD radiation is
investigated by means of the subjet topology, providing a more
stringent test of the pQCD calculations.

\begin{figure}[p]
\vfill
\setlength{\unitlength}{1.0cm}
\begin{picture} (18.0,10.0)
\put (0.0,0.0){\centerline{\epsfig{figure=\figdir 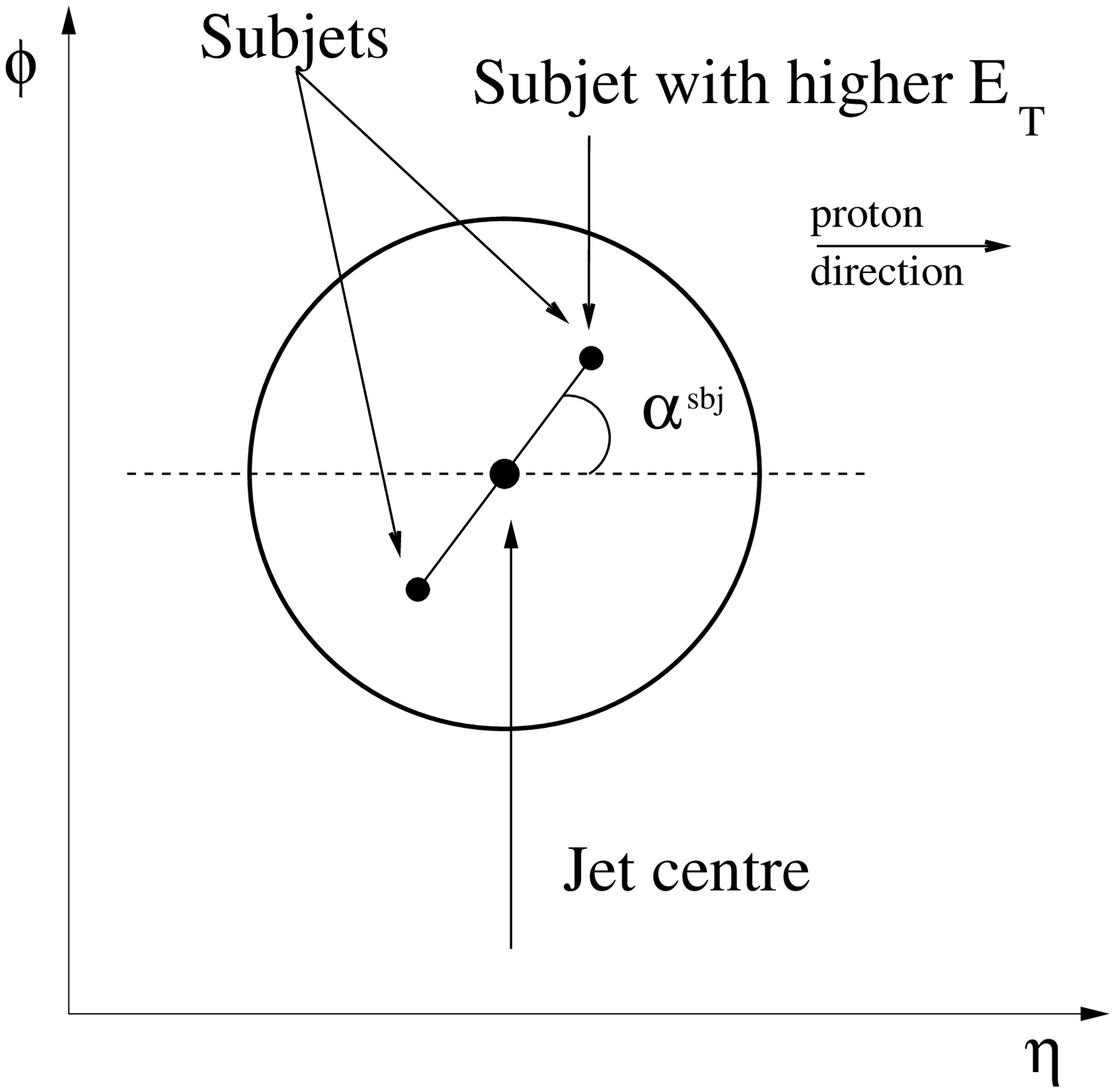,width=10cm}}}
\end{picture}
\caption
{\it 
Schematic representation of the $\asbj$ variable.
}
\label{fig0}
\vfill
\end{figure}

In this paper, measurements of normalised differential subjet cross
sections for those jets which contain two subjets at a given
resolution scale are presented. The
measurements were done as functions of the ratio between the subjet
transverse energy and that of the jet, $\etsbj/\etjet$, the difference
between the subjet pseudorapidity\footnote{The ZEUS coordinate system
  is a right-handed Cartesian system, with the $Z$ axis pointing in
  the proton beam direction, referred to as the ``forward direction'',
  and the $X$ axis pointing left towards the centre of HERA. The
  coordinate origin is at the nominal interaction point. The
  pseudorapidity is defined as $\eta=-\ln(\tan\frac{\theta}{2})$,
  where the polar angle $\theta$ is taken with respect to the proton
  beam direction.} (azimuth) and that of the jet,
$\etasbj-\etajet$ ($|\phisbj-\phijet|$), and $\asbj$, the angle, as
viewed from the jet centre, between the subjet with higher transverse
energy and the proton beam line in the pseudorapidity-azimuth plane (see
Fig.~\ref{fig0}). The predictions of pQCD at next-to-leading order
(NLO) were compared to the data.

\section{Jets and subjets}
\label{jssmdef}

The analysis of subjets presented in this paper was performed using
the laboratory frame. In this frame, the calculations of the subjet
distributions can be performed up to $\oass$, i.e. NLO, with jets
consisting of up to three partons. The analysis used events with
high virtuality of the exchanged boson, $\q2$; at low values of $\q2$,
the sample of events with at least one jet of high $\etjet$
($\etjet\gg\sqrt{\q2}$) is dominated by dijet events. In that case,
the calculations include jets consisting of up to only two partons
and, therefore, correspond to lowest-order predictions of jet
substructure.

The $\kt$ cluster algorithm~\cite{np:b406:187} was used in the
longitudinally invariant inclusive mode~\cite{pr:d48:3160} to define
jets in the hadronic final state. Subjets~\cite{np:b383:419,*np:b421:545,*pl:b378:279,*jhep:9909:009} were resolved within a jet
by considering all particles associated with the jet and repeating the
application of the $\kt$ cluster algorithm until, for every pair of
particles $i$ and $j$ the quantity 
$d_{ij}={\rm min}(E_{T,i},E_{T,j})^2\cdot((\eta_i-\eta_j)^2+(\phi_i-\phi_j)^2)$, 
where $E_{T,i}$, $\eta_i$ and $\phi_i$ are the transverse energy,
pseudorapidity and azimuth of particle $i$, respectively,
was greater than $d_{\rm cut}=\yc\cdot(\etjet)^2$. All remaining
clusters were called subjets. 

The subjet multiplicity depends upon the value chosen for the
resolution parameter $\yc$. Subjet distributions were studied for
those jets with exactly two subjets at a value of the resolution
parameter of $\yc=0.05$. This value of $\yc$ was chosen as a
compromise between resolution, size of the hadronisation correction
factors and statistics. The effect of the parton-to-hadron corrections
on the shape of the subjet distributions becomes increasingly larger
as $\yc$ decreases. On the other hand, the number of jets with exactly
two subjets decreases rapidly as $\yc$ increases.

Subjet distributions were studied as functions of $\etsbj/\etjet$, 
$\etasbj-\etajet$, $|\phisbj-\phijet|$ and $\asbj$. One of the goals
of this study was to investigate the extent to which pQCD
calculations are able to reproduce the observed distributions. In
addition, the dependence of the splitting functions $P_{ab}(z,\mu)$ on
$z$ can be investigated using the $\etsbj/\etjet$ distribution. The
splitting functions at leading order (LO) do not depend on $\mu$ but
acquire a weak dependence due to higher-order corrections. Such a
dependence can be investigated by measuring the subjet distributions
in different regions of $\etjet$ or $\q2$.

The substructure of jets consisting of a quark-gluon pair
(the quark-induced process $eq\rightarrow eqg$) or a quark-antiquark
pair (the gluon-induced process $eg\rightarrow e\qq$) are predicted to
be different (see Section 8.1). Furthermore, the relative
contributions of quark- and gluon-induced processes vary with
Bjorken $x$ and $\q2$. The predicted difference mentioned
above is amenable to experimental investigation by comparing the shape
of the subjet distributions in different regions of $x$ and $\q2$.

Colour coherence leads to a suppression of soft-gluon radiation in
certain regions of phase space. The effects of colour coherence
between the initial and final states have been studied in
hadron-hadron collisions~\cite{pr:d50:5562}. These effects are also
expected to appear in lepton-hadron collisions. For the process
$eq\rightarrow eqg$, colour coherence implies a tendency of the subjet
with lower (higher) transverse energy, $E_{T,{\rm low}}^{\rm sbj}$
($E_{T,{\rm high}}^{\rm sbj}$), to have $\etasbj-\etajet>0$
($\etasbj-\etajet<0$). The variable $\asbj$, defined in close analogy
to the variables used to study colour coherence in hadron-hadron
collisions~\cite{pr:d50:5562}, reflects directly whether the subjet
with the lower transverse energy has a tendency to be emitted towards
the proton beam direction.

\section{Experimental set-up}

A detailed description of the ZEUS detector can be found
elsewhere~\cite{pl:b293:465,zeus:1993:bluebook}. A brief outline of
the components most relevant for this analysis is given below.

Charged particles were tracked in the central tracking detector
(CTD)~\cite{nim:a279:290,*npps:b32:181,*nim:a338:254}, which operated
in a magnetic field of $1.43\Tesla$ provided by a thin superconducting
solenoid. The CTD consisted of $72$~cylindrical drift-chamber
layers, organised in nine superlayers covering the
polar-angle
region \mbox{$15^\circ<\theta<164^\circ$}. The transverse-momentum
resolution for full-length tracks can be parameterised as
$\sigma(p_T)/p_T=0.0058p_T\oplus0.0065\oplus0.0014/p_T$, with $p_T$ in
$\Gev$. The tracking system was used to measure the interaction vertex
with a typical resolution along (transverse to) the beam direction of
$0.4$~($0.1$)~cm and to cross-check the energy scale of the calorimeter.

The high-resolution uranium--scintillator calorimeter
(CAL)~\cite{nim:a309:77,*nim:a309:101,*nim:a321:356,*nim:a336:23} covered
$99.7\%$ of the total solid angle and consisted of three parts: the
forward (FCAL), the barrel (BCAL) and the rear (RCAL)
calorimeters. Each part was subdivided transversely into towers and 
longitudinally into one electromagnetic section and either one
(in RCAL) or two (in BCAL and FCAL) hadronic sections. The
smallest subdivision of the calorimeter was called a cell. Under
test-beam conditions, the CAL single-particle relative energy
resolutions were $\sigma(E)/E=0.18/\sqrt E$ for electrons and
$\sigma(E)/E=0.35/\sqrt E$ for hadrons, with $E$ in GeV.

The luminosity was measured from the rate of the bremsstrahlung process
$ep\rightarrow e\gamma p$. The resulting small-angle energetic photons
were measured by the luminosity
monitor~\cite{desy-92-066,*zfp:c63:391,*acpp:b32:2025}, a
lead--scintillator calorimeter placed in the HERA tunnel at $Z=-107$ m.

\section{Data selection}

The data were collected during the running period 1998--2000, when HERA
operated with protons of energy $E_p=920$~GeV and electrons or
positrons\footnote{In the following, the term ``electron''
  denotes generically both the electron ($e^-$) and the positron
  ($e^+$).} of energy $E_e=27.5$~GeV, and correspond to an integrated
luminosity of $81.7\pm 1.9$~\pb1.

Neutral current DIS events were selected offline using criteria
similar to those reported previously~\cite{np:b700:3}. The main
steps are given below.

A reconstructed event vertex consistent with the nominal interaction
position was required and cuts based on tracking information were
applied to reduce the contamination from beam-induced and cosmic-ray
background. The scattered-electron candidate was identified using the
pattern of energy deposits in the
CAL~\cite{nim:a365:508,*nim:a391:360}. The energy, $E_e^{\prime}$, and
polar angle, $\theta_e$, of the electron candidate were also
determined from the CAL measurements. The double-angle
method~\cite{proc:hera:1991:23,*proc:hera:1991:43}, which uses
$\theta_e$ and an angle $\gamma$ that corresponds, in the quark-parton
model, to the direction of the scattered quark, was used to
reconstruct $\q2$. The angle $\gamma$ was reconstructed using the CAL
measurements of the hadronic final state.

Electron candidates were required to have an energy
$E_e^{\prime}>10$~GeV, to ensure a high and well understood
electron-finding efficiency and to suppress background from
photoproduction. The inelasticity variable, $y$, as reconstructed
using the electron energy and polar angle, was required to be below
$0.95$; this condition removed events in which fake electron
candidates from photoproduction background were found in the FCAL. The
requirement $38<(E-p_Z)<65$~GeV, where $E$ is the total CAL energy and
$p_Z$ is the $Z$ component of the energy measured in the CAL cells,
was applied to remove events with large initial-state radiation and to
reduce further the photoproduction background. Remaining cosmic rays
and beam-related background were rejected by requiring the total
missing transverse momentum, $\ptmis$, to be small compared to the
total transverse energy, $E^{\rm tot}_T$, 
\mbox{$\ptmis/\sqrt{E^{\rm tot}_T}<3\ \sqrt{\rm GeV}$}. The kinematic
range was restricted to $\q2>125$ \g2.

The $\kt$ cluster algorithm was used in the longitudinally invariant
inclusive mode to reconstruct jets in the measured hadronic final
state from the energy deposits in the CAL cells. The jet algorithm was
applied after excluding those cells associated with the
scattered-electron candidate. Jet transverse-energy corrections were
computed using the method developed in a previous
analysis~\cite{np:b700:3}. Events were required to have at least one
jet of $\etjet>14$~GeV and $\etar$. The final sample of 128986 events
contained 132818 jets, of which 21162 jets had exactly two subjets at
$\yc=0.05$.

\section{Monte Carlo simulation}
\label{mc}

Samples of events were generated to determine the response of the
detector to jets of hadrons and the correction factors necessary to
obtain the hadron-level subjet cross sections. The hadron level is
defined as those hadrons with lifetime $\tau\geq 10$~ps. The generated
events were passed through the {\sc
  Geant}~3.13-based~\cite{tech:cern-dd-ee-84-1} ZEUS detector- and 
trigger-simulation programs \cite{zeus:1993:bluebook}. They were
reconstructed and analysed applying the same program chain as to the
data.
 
Neutral current DIS events including radiative effects were simulated
using the {\sc Heracles}~4.6.1~\cite{cpc:69:155,*spi:www:heracles}
program with the {\sc
  Djangoh}~1.1~\cite{cpc:81:381,*spi:www:djangoh11} interface to the
hadronisation programs. {\sc Heracles} includes corrections for
initial- and final-state radiation, vertex and propagator terms, and
two-boson exchange. The QCD cascade is simulated using the
colour-dipole
model (CDM)~\cite{pl:b165:147,*pl:b175:453,*np:b306:746,*zfp:c43:625}
including the LO QCD diagrams as implemented in
{\sc Ariadne}~4.08~\cite{cpc:71:15,*zfp:c65:285} and, alternatively,
with the MEPS model of {\sc Lepto} 6.5~\cite{cpc:101:108}. The
CTEQ5D~\cite{epj:c12:375} proton PDFs were used for these
simulations. Fragmentation into hadrons is performed using the Lund
string model~\cite{prep:97:31} as implemented in 
{\sc Jetset}~\cite{cpc:82:74,*cpc:135:238,cpc:39:347,*cpc:43:367}.
 
The jet search was performed on the Monte Carlo (MC) events using the
energy measured in the CAL cells in the same way as for the data. The
same jet algorithm was also applied to the final-state particles
(hadron level) and to the partons available after the parton shower
(parton level) to compute hadronisation correction factors (see
Section~\ref{nlo}).

\section{QCD calculations}
\label{nlo}

The $\oass$ NLO QCD calculations used to compare with the data
are based on the program {\sc Disent}~\cite{np:b485:291}. The
calculations used a generalised version of the subtraction
method~\cite{np:b178:421} and were performed in the massless
$\overline{\rm MS}$ renormalisation and factorisation schemes. The
number of flavours was set to five; the renormalisation ($\mu_R$) and
factorisation ($\mu_F$) scales were set to $\mu_R=\mu_F=Q$; $\as$
was calculated at two loops using 
$\Lambda^{(5)}_{\overline{\rm MS}}=220$~MeV which corresponds to 
$\asz=0.118$. The ZEUS-S~\cite{pr:d67:012007} parameterisations of the
proton PDFs were used. The results obtained with {\sc Disent} were
cross-checked by using the program {\sc Nlojet++}~\cite{prl:87:082001}.

Since the measurements refer to jets of hadrons, whereas the QCD
calculations refer to jets of partons, the predictions were corrected
to the hadron level using the MC samples described in
Section~\ref{mc}. The multiplicative correction factor, $C_{\rm had}$,
defined as the ratio of the cross section for subjets of
hadrons to that of partons, was estimated with the {\sc Lepto}-MEPS
model, since it reproduced the shape of the QCD calculations better. 
The normalised cross-section calculations changed typically by less
than $\pm 20\%$ upon application of the parton-to-hadron corrections,
except at the edges of the distributions, where they changed by up to
$\pm 50\%$. Other effects not accounted for in the calculations,
namely QED radiative corrections and $\z0$ exchange, were found to be
very small for the normalised cross-section calculations and
neglected.

The following theoretical uncertainties were considered (as
examples of the size of the uncertainties, average values of the
effect of each uncertainty on the normalised cross section as
functions of $\etsbj/\etjet$, $\etasbj-\etajet$, $|\phisbj-\phijet|$
and $\asbj$ are given in parentheses):
\begin{itemize}
  \item the uncertainty in the modelling of the parton shower was
    estimated by using different models (see Section~\ref{mc}) to
    calculate the parton-to-hadron correction factors
    ($5.6\%$, $13.2\%$, $7.6\%$, $5.3\%$);
  \item the uncertainty on the calculations due to higher-order terms
    was estimated by varying $\mu_R$ by a factor of two up and down 
    ($0.01\%$, $0.46\%$, $0.58\%$, $0.34\%$);
  \item the uncertainty on the calculations due to the choice of
    $\mu_F$ was estimated by varying $\mu_F$ by a factor of two up and
    down
    ($0.05\%$, $0.43\%$, $0.11\%$, $0.12\%$);
  \item the uncertainty on the calculations due to those on the proton
    PDFs was estimated by repeating the calculations using 22
    additional sets from the ZEUS analysis~\cite{pr:d67:012007}; this
    analysis takes into account the statistical and correlated
    systematic experimental uncertainties of each data set used in the
    determination of the proton PDFs
    ($0.07\%$, $0.18\%$, $0.12\%$, $0.05\%$);
  \item the uncertainty on the calculations due to that on $\asz$ was
    estimated by repeating the calculations using two additional sets
    of proton PDFs, for which different values of $\asz$ were assumed
    in the fits. The difference between the calculations using these
    various sets was scaled to reflect the uncertainty on the current
    world average of $\as$~\cite{jp:g26:r27}
    ($0.02\%$, $0.04\%$, $0.05\%$, $0.01\%$).
\end{itemize}

These uncertainties were added in quadrature and are shown as
hatched bands in the figures.

\section{Corrections and systematic uncertainties}

\begin{figure}[p]
\vfill
\setlength{\unitlength}{1.0cm}
\begin{picture} (18.0,17.0)
\put (0.5,0.0){\centerline{\epsfig{figure=\figdir 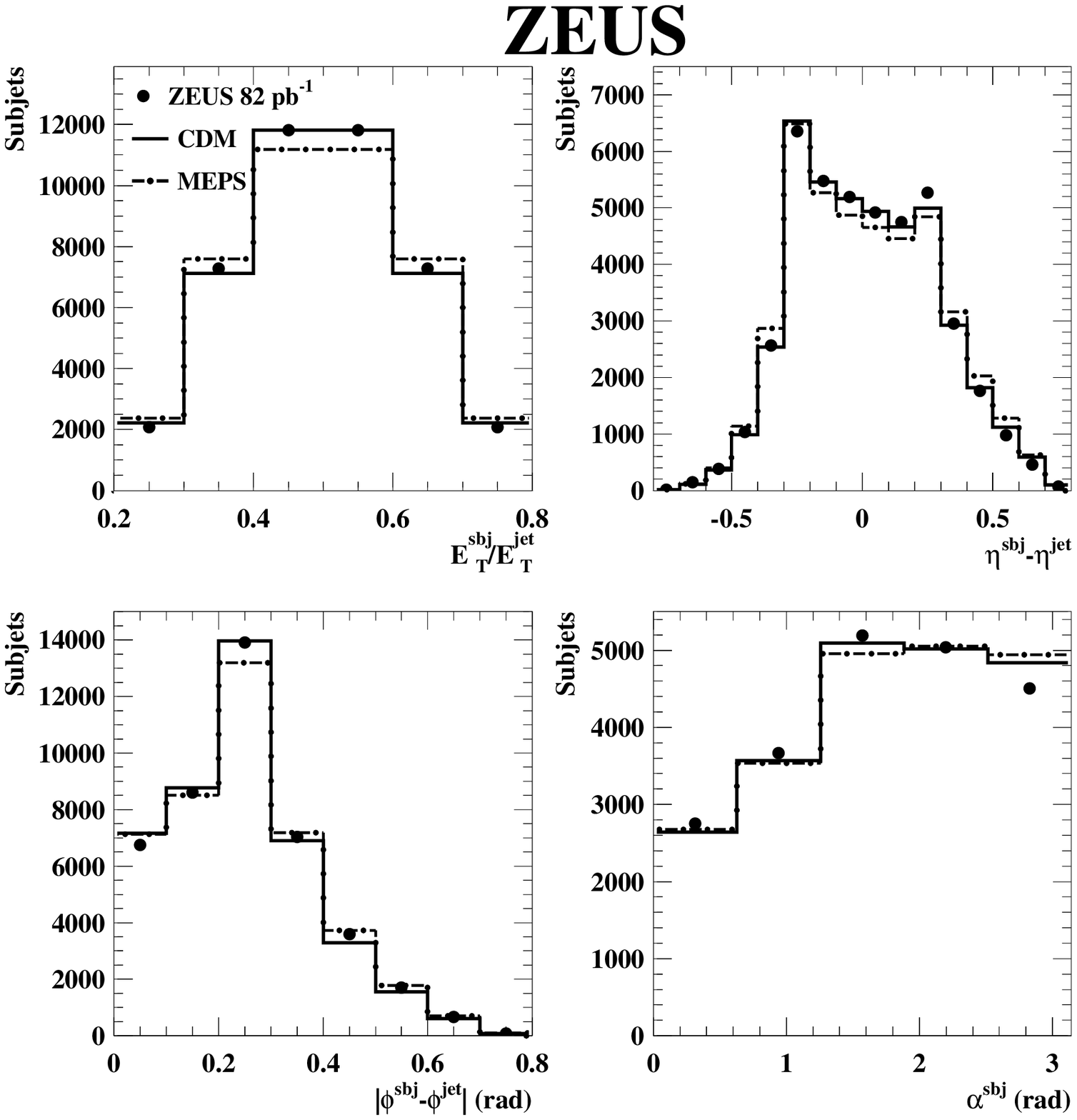,width=20cm}}}
\put (6.5,17.4){\bf\small (a)}
\put (10.1,17.4){\bf\small (b)}
\put (6.5,8.4){\bf\small (c)}
\put (10.1,8.4){\bf\small (d)}
\end{picture}
\caption
{\it 
Detector-level normalised subjet data distributions (dots) for jets
with $\etjet>14$~GeV and $-1<\etajet<2.5$ which have two
subjets for $\yc=0.05$ in the kinematic region given by $\q2>125$~\gev$^2$
as functions of (a) $\etsbj/\etjet$, (b) $\etasbj-\etajet$, (c)
$|\phisbj-\phijet|$ and (d) $\asbj$. The statistical uncertainties are
smaller than the marker size. For comparison, the
distributions of the CDM (solid histograms) and MEPS (dot-dashed
histograms) Monte Carlo models are included.
}
\label{fig1}
\vfill
\end{figure}

The sample of events generated with CDM, after applying the same
offline selection as for the data, gives a reasonably good description
of the measured distributions of the kinematic, jet and subjet
variables; the description provided by the MEPS sample is somewhat
poorer. The comparison of the measured subjet distributions and the MC
simulations is shown in Fig.~\ref{fig1}. 

The normalised differential cross sections were obtained from the data
using the bin-by-bin correction method,

$$\frac{1}{\sigma}\frac{d\sigma_i}{dA}=\frac{1}{\sigma}\frac{N_{{\rm data},i}}{{\cal L}\cdot \Delta A_i}\cdot\frac{N^{\rm had}_{{\rm MC},i}}{N^{\rm det}_{{\rm MC},i}},$$
where $N_{{\rm data},i}$ is the number of subjets in data in bin $i$
of the subjet variable $A$, 
$N^{\rm had}_{{\rm MC},i}\ (N^{\rm det}_{{\rm MC},i})$ is the number
of subjets in MC at hadron (detector) level, ${\cal L}$ is the
integrated luminosity and $\Delta A_i$ is the bin width.
The MC samples of CDM and MEPS were used to compute the acceptance
correction factors to the subjet distributions. These correction
factors took into account the efficiency of the trigger, the selection
criteria and the purity and efficiency of the jet and subjet
reconstruction.

The following sources of systematic uncertainty were
considered for the measured subjet cross sections (as
examples of the size of the uncertainties, average values of the
effect of each uncertainty on the normalised cross section as
functions of $\etsbj/\etjet$, $\etasbj-\etajet$, $|\phisbj-\phijet|$
and $\asbj$ are given in parentheses):
\begin{itemize}
  \item the deviations in the results obtained by using either
    CDM or MEPS to correct the data from their average were taken to
    represent systematic uncertainties due to the modelling of the
    parton shower
    ($0.5\%$, $2.9\%$, $2.6\%$, $1.3\%$);
  \item variations in the simulation of the CAL response to low-energy
    particles
    ($0.3\%$, $1.6\%$, $1.2\%$, $0.6\%$).
\end{itemize}

Other uncertainties, such as those arising from the uncertainty in the
absolute energy scale of the
jets~\cite{pl:b531:9,*epj:c23:615,proc:calor:2002:767}, the
uncertainty in the simulation of the trigger and the uncertainty in
the absolute energy scale of the electron
candidate~\cite{epj:c21:443}, were investigated and found to be
negligible. The systematic uncertainties were added in quadrature to
the statistical uncertainties and are shown as error bars in the
figures.

\section{Results}
\label{results}

Normalised differential subjet cross sections were measured for
$\q2>125$ \g2\ for jets with $\etjet>14$ GeV and $-1<\etajet<2.5$
which have exactly two subjets for $\yc=0.05$.

\begin{figure}[p]
\vfill
\setlength{\unitlength}{1.0cm}
\begin{picture} (18.0,15.0)
\put (0.0,0.0){\centerline{\epsfig{figure=\figdir 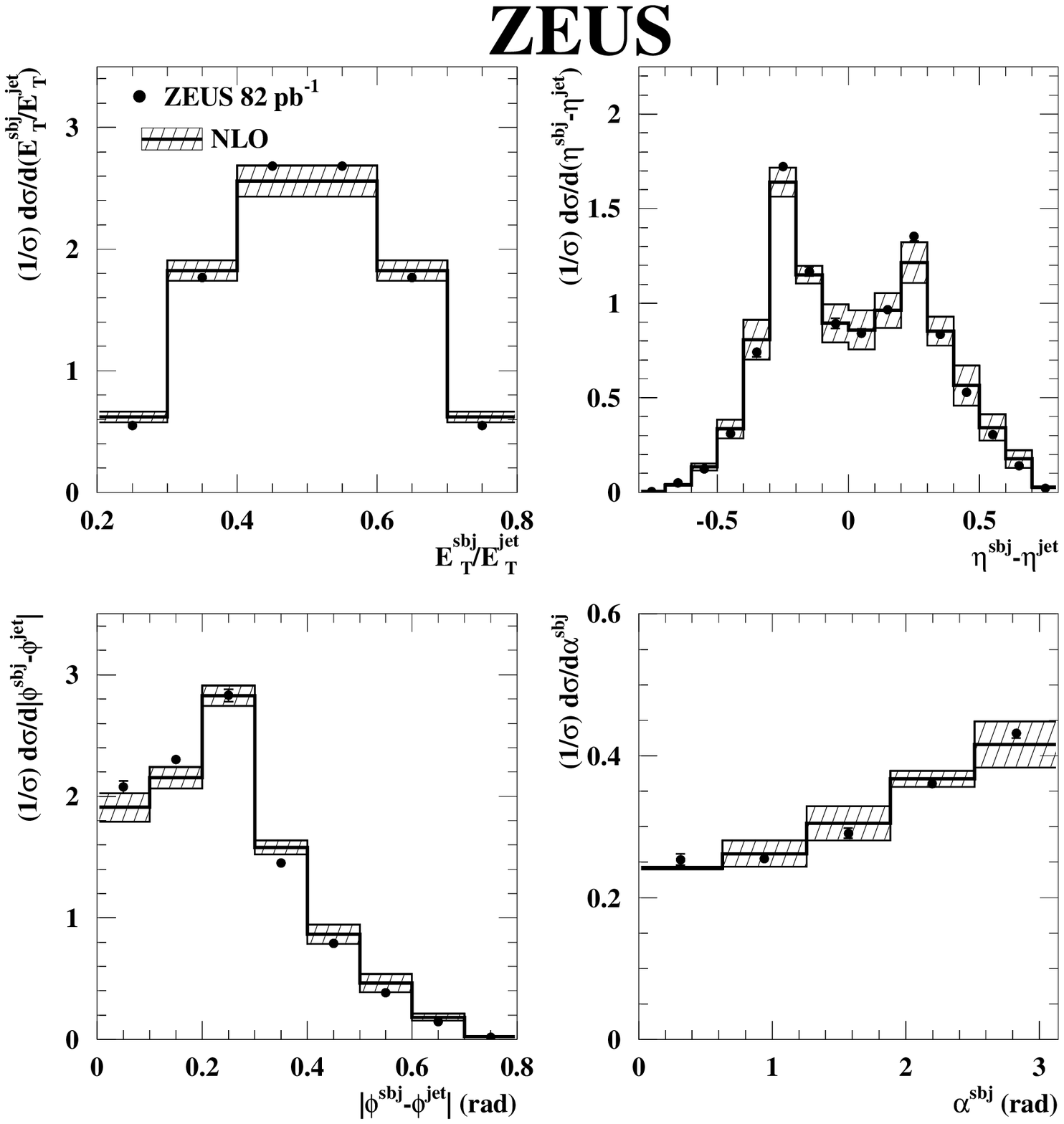,width=20cm}}}
\put (6.0,17.4){\bf\small (a)}
\put (9.6,17.4){\bf\small (b)}
\put (6.0,8.4){\bf\small (c)}
\put (9.6,8.4){\bf\small (d)}
\end{picture}
\caption
{\it 
Measured normalised differential subjet cross sections (dots) for jets
with $\etjet>14$ GeV and $-1<\etajet<2.5$ which have two subjets for
$\yc=0.05$ in the kinematic region given by $\q2>125$~\gev$^2$ as
functions of (a) $\etsbj/\etjet$, (b) $\etasbj-\etajet$, 
(c) $|\phisbj-\phijet|$ and (d) $\asbj$. The inner error bars
represent the statistical uncertainties of the data, the outer
error bars show the statistical and systematic uncertainties added in
quadrature. In many cases, the error bars are smaller than the marker
size and are therefore not visible.
For comparison, the NLO QCD predictions (solid histograms)
are included. The hatched bands represent the theoretical
uncertainty.
}
\label{fig2}
\vfill
\end{figure}

The distribution of the fraction of transverse energy,
$(1/\sigma)(d\sigma/d(\etsbj/\etjet))$, is presented in
Fig.~\ref{fig2}a. It contains two entries per jet and is symmetric
with respect to $\etsbj/\etjet=0.5$ by construction. This distribution
has a peak for $0.4<\etsbj/\etjet<0.6$, which shows that the two
subjets tend to have similar transverse energies.

The $\etasbj-\etajet$ data distribution is shown in Fig.~\ref{fig2}b
and also has two entries per jet. The measured cross section has a
two-peak structure; the dip around $\etasbj-\etajet=0$ is due to the
fact that the two subjets are not resolved when they are too close
together.

Figure~\ref{fig2}c presents the measured normalised cross section as a
function of $|\phisbj-\phijet|$. There are two entries per jet in this
distribution. The distribution has a peak for
$0.2<|\phisbj-\phijet|<0.3$; the suppression around
$|\phisbj-\phijet|=0$ also arises from the fact that the two subjets
are not resolved when they are too close together.

\begin{figure}[p]
\vfill
\setlength{\unitlength}{1.0cm}
\begin{picture} (18.0,10.0)
\put (0.0,-2.0){\centerline{\epsfig{figure=\figdir 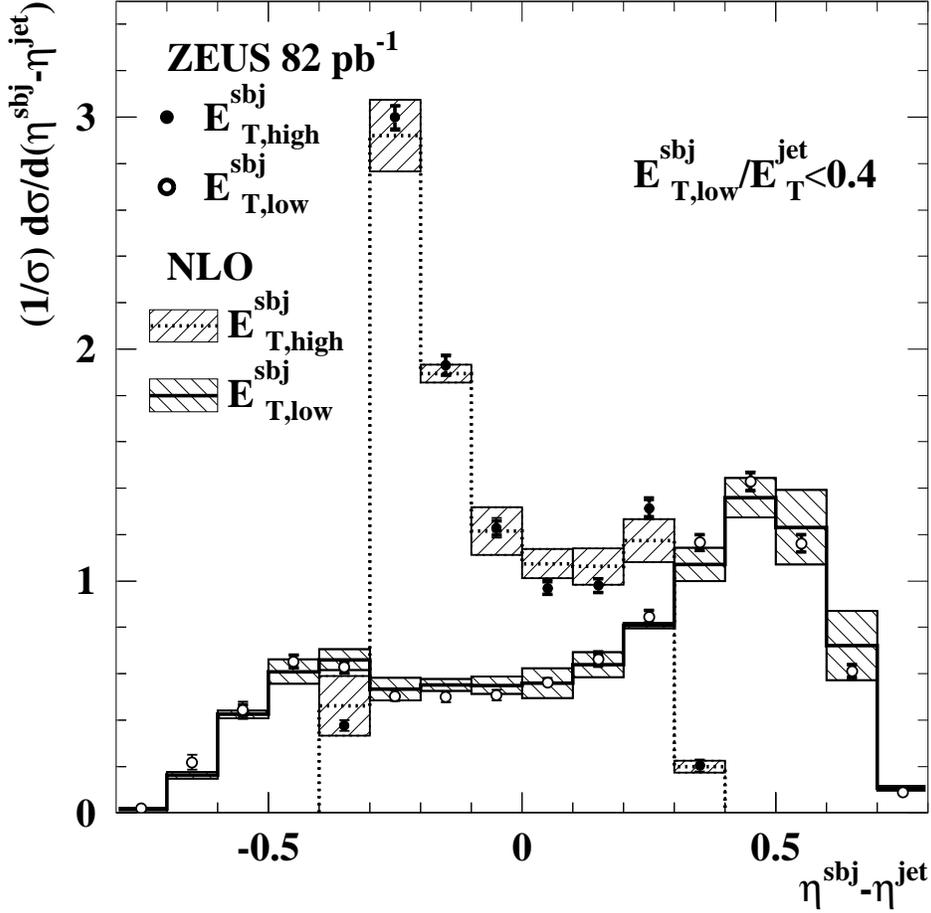,width=18cm}}}
\end{picture}
\caption
{\it 
Measured normalised differential subjet cross sections for jets 
with $\etjet>14$ GeV and $-1<\etajet<2.5$ which have two
subjets for $\yc=0.05$ in the kinematic region given by $\q2>125$~\gev$^2$
and $E_{T,{\rm low}}^{\rm sbj}/\etjet<0.4$ as functions of
$\etasbj-\etajet$ separately for the higher (dots) and lower (open
circles) $\etsbj$ subjets. Other details are as in the caption to
Fig.~\ref{fig2}.
}
\label{fig3}
\vfill
\end{figure}

The data distribution as a function of $\asbj$ (one entry per jet)
increases as $\asbj$ increases (see Fig.~\ref{fig2}d). This shows that
the subjet with higher transverse energy tends to be in the rear
direction. This is consistent with the asymmetric peaks observed in
the $\etasbj-\etajet$ distribution (see
Fig.~\ref{fig2}b). Figure~\ref{fig3} shows the 
$\etasbj-\etajet$ distribution for those jets which have two subjets
with asymmetric $\etsbj$ ($E_{T,{\rm low}}^{\rm sbj}/\etjet<0.4$,
or, equivalently, $E_{T,{\rm high}}^{\rm sbj}/\etjet>0.6$),
separately for the subjet with higher and lower $\etsbj$. It is
to be noted that since the jet axis is reconstructed as the
transverse-energy-weighted average of the subjet axes, the subjet with
higher $\etsbj$ is constrained to be closer to the jet axis than that
of the lower $\etsbj$ subjet. The measured distributions show
that the higher (lower) $\etsbj$ subjet tends to be in the rear
(forward) direction. All these observations support the expectation of
the presence of colour-coherence effects between the initial and final
states and, in particular, the tendency of the subjet with lower
$\etsbj$ to be emitted predominantly towards the proton beam
direction.

\subsection{Comparison with NLO QCD calculations}

Next-to-leading-order QCD calculations are compared to the data in
Figs.~\ref{fig2} and \ref{fig3}. The QCD predictions give an adequate
description of the data. However, the data points are situated at the
upper (lower) edge of the theoretical uncertainty in some regions of
the subjet variables such as $\etsbj/\etjet\sim 0.5$,
$|\phisbj-\phijet|\sim 0$, $\asbj\sim 0$ and the peaks in the
$\etasbj-\etajet$ distribution ($\etsbj/\etjet\sim 0.25$,
$|\phisbj-\phijet|>0.3$ and $|\etasbj-\etajet|>0.5$). Since the
calculations are normalised to unity, the uncertainties are correlated
among the points; this correlation gives rise to the
pulsating pattern exhibited by the theoretical uncertainties.

The calculation of the cross section as a function of $\etsbj/\etjet$
exhibits a peak at $0.4<\etsbj/\etjet<0.6$, as seen in the data. The
calculations for the $\etasbj-\etajet$ and $\asbj$ distributions
predict that the subjet with higher transverse energy tends to be in the
rear direction, in agreement with the data. This shows that the
mechanism driving the subjet topology in the data is the
$eq\rightarrow eqg$ and $eg\rightarrow e\qq$ subprocesses as
implemented in the pQCD calculations.

\begin{figure}[p]
\vfill
\setlength{\unitlength}{1.0cm}
\begin{picture} (18.0,17.0)
\put (0.0,0.0){\centerline{\epsfig{figure=\figdir 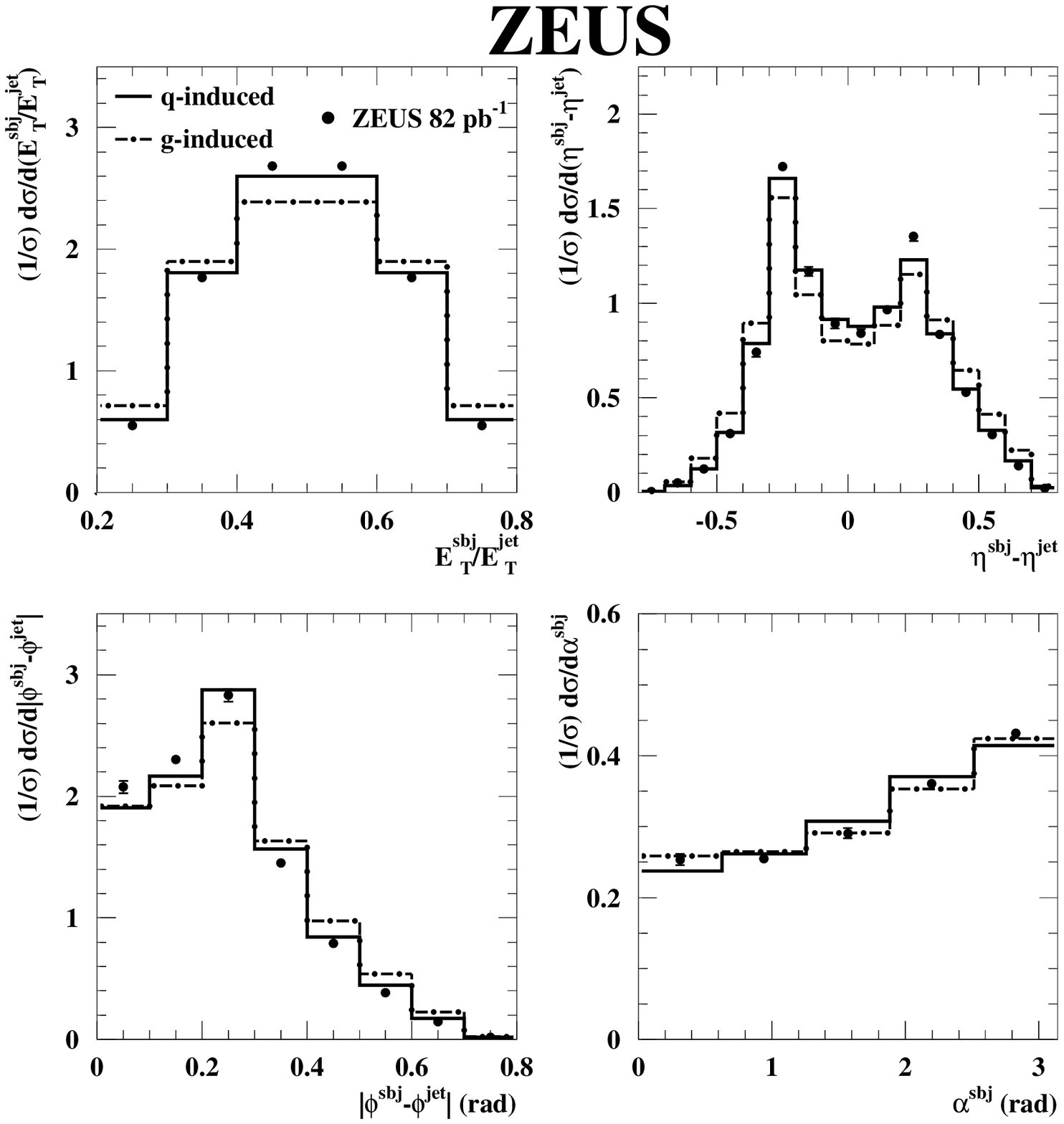,width=20cm}}}
\put (6.0,17.4){\bf\small (a)}
\put (9.6,17.4){\bf\small (b)}
\put (6.0,8.4){\bf\small (c)}
\put (9.6,8.4){\bf\small (d)}
\end{picture}
\caption
{\it 
Measured normalised differential subjet cross sections (dots) for jets
with $\etjet>14$ GeV and $-1<\etajet<2.5$ which have two subjets for
$\yc=0.05$ in the kinematic region given by $\q2>125$~\gev$^2$ as
functions of (a) $\etsbj/\etjet$, (b) $\etasbj-\etajet$, (c)
$|\phisbj-\phijet|$ and (d) $\asbj$. For comparison, the NLO
predictions for quark- (solid histograms) and gluon-induced (dot-dashed
histograms) processes are included. Other details are as in the
caption to Fig.~\ref{fig2}.
}
\label{fig4}
\vfill
\end{figure}

To gain further insight into the pattern of parton radiation, the
predictions for quark- and gluon-induced processes (see Section 2) are
compared separately with the data in Fig.~\ref{fig4}. The NLO
calculations predict that the two-subjet rate is dominated by
quark-induced processes: the relative contribution of quark- (gluon-)
induced processes is $81\%$ ($19\%$). The shape of the predictions for
these two types of processes are different; in quark-induced
processes, the two subjets have more similar transverse energies (see
Fig.~\ref{fig4}a) and are closer to each other (see Fig.~\ref{fig4}b
and \ref{fig4}c) than in gluon-induced processes. The comparison with
the measurements shows that the data are better described by the
calculations for jets arising from a $qg$ pair than those coming from
a $\qq$ pair. 

\subsection{{\boldmath $\etjet$}, {\boldmath $\q2$} and {\boldmath
    $x$} dependence of the subjet distributions}

\begin{figure}[p]
\vfill
\setlength{\unitlength}{1.0cm}
\begin{picture} (18.0,15.0)
\put (0.0,0.0){\centerline{\epsfig{figure=\figdir 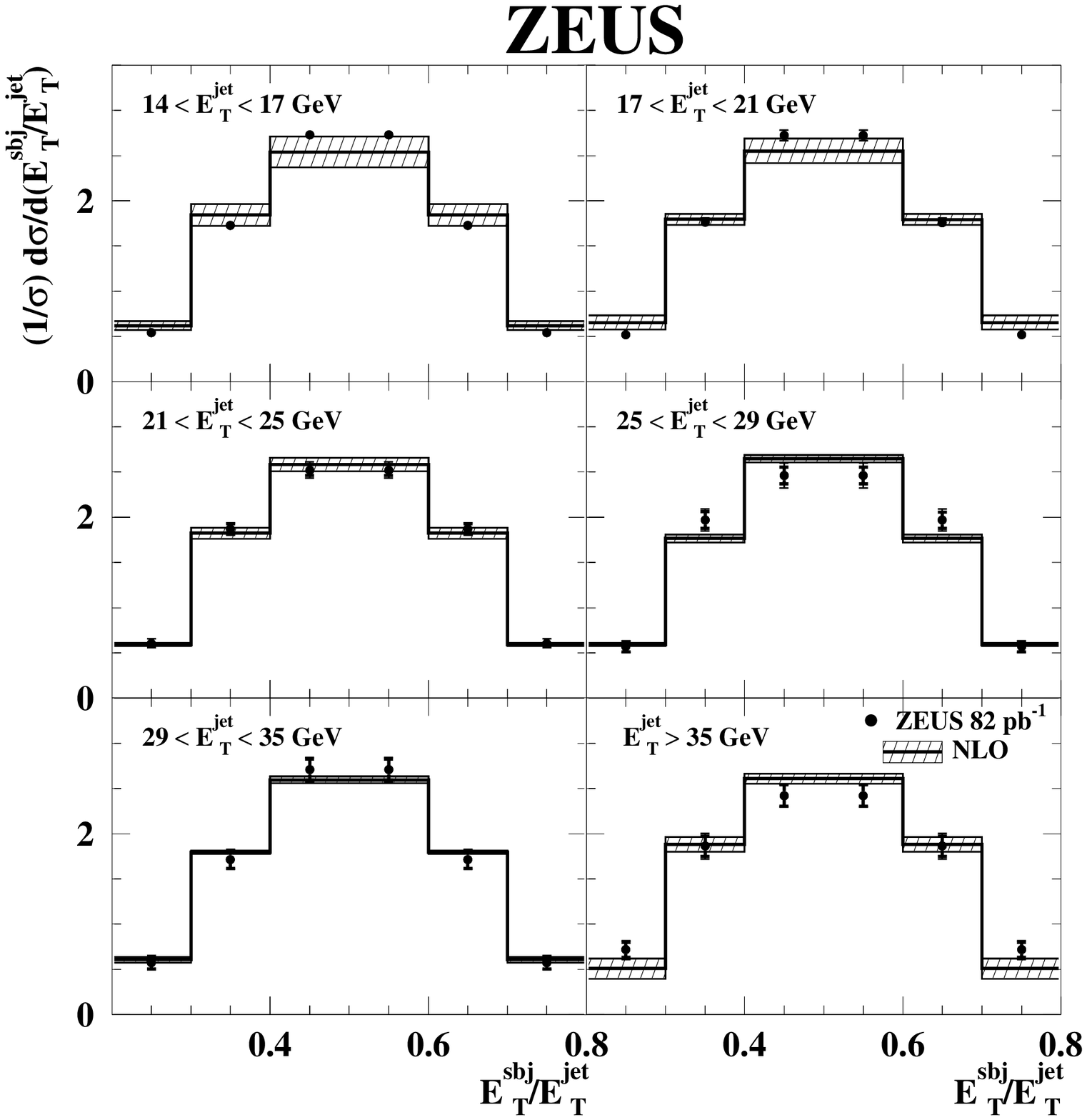,width=18cm}}}
\end{picture}
\caption
{\it 
Measured normalised differential subjet cross sections (dots) for jets
with $\etjet>14$ GeV and $-1<\etajet<2.5$ which have two subjets for
$\yc=0.05$ in the kinematic region given by $\q2>125$~\gev$^2$
as functions of $\etsbj/\etjet$ in different regions of
$\etjet$. Other details are as in the caption to
Fig.~\ref{fig2}.
}
\label{fig5}
\vfill
\end{figure}

\begin{figure}[p]
\vfill
\setlength{\unitlength}{1.0cm}
\begin{picture} (18.0,15.0)
\put (0.0,0.0){\centerline{\epsfig{figure=\figdir 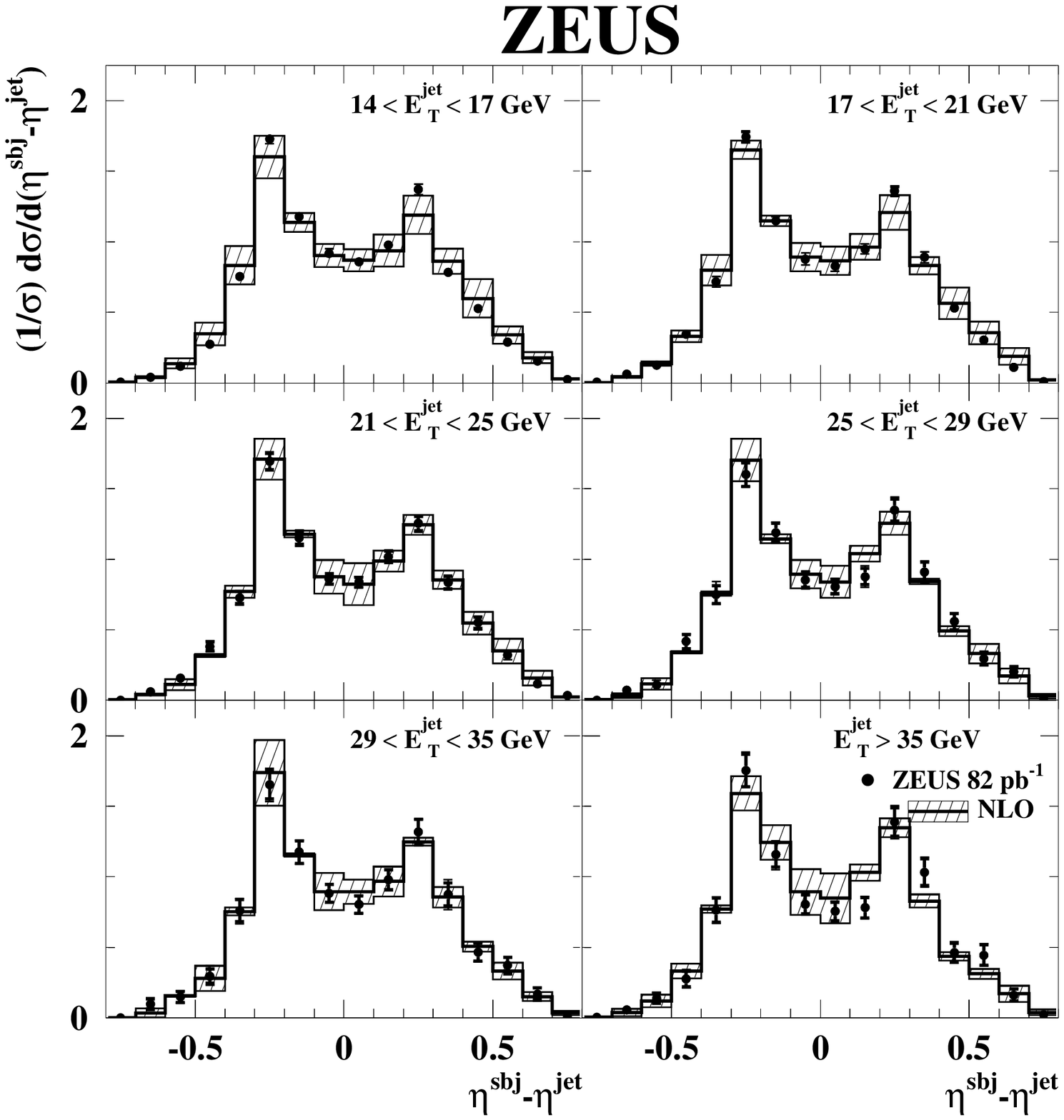,width=18cm}}}
\end{picture}
\caption
{\it 
Measured normalised differential subjet cross sections (dots) for jets
with $\etjet>14$ GeV and $-1<\etajet<2.5$ which have two subjets for
$\yc=0.05$ in the kinematic region given by $\q2>125$~\gev$^2$
as functions of $\etasbj-\etajet$ in different regions of
$\etjet$. Other details are as in the caption to Fig.~\ref{fig2}.
}
\label{fig6}
\vfill
\end{figure}

\begin{figure}[p]
\vfill
\setlength{\unitlength}{1.0cm}
\begin{picture} (18.0,15.0)
\put (0.0,0.0){\centerline{\epsfig{figure=\figdir 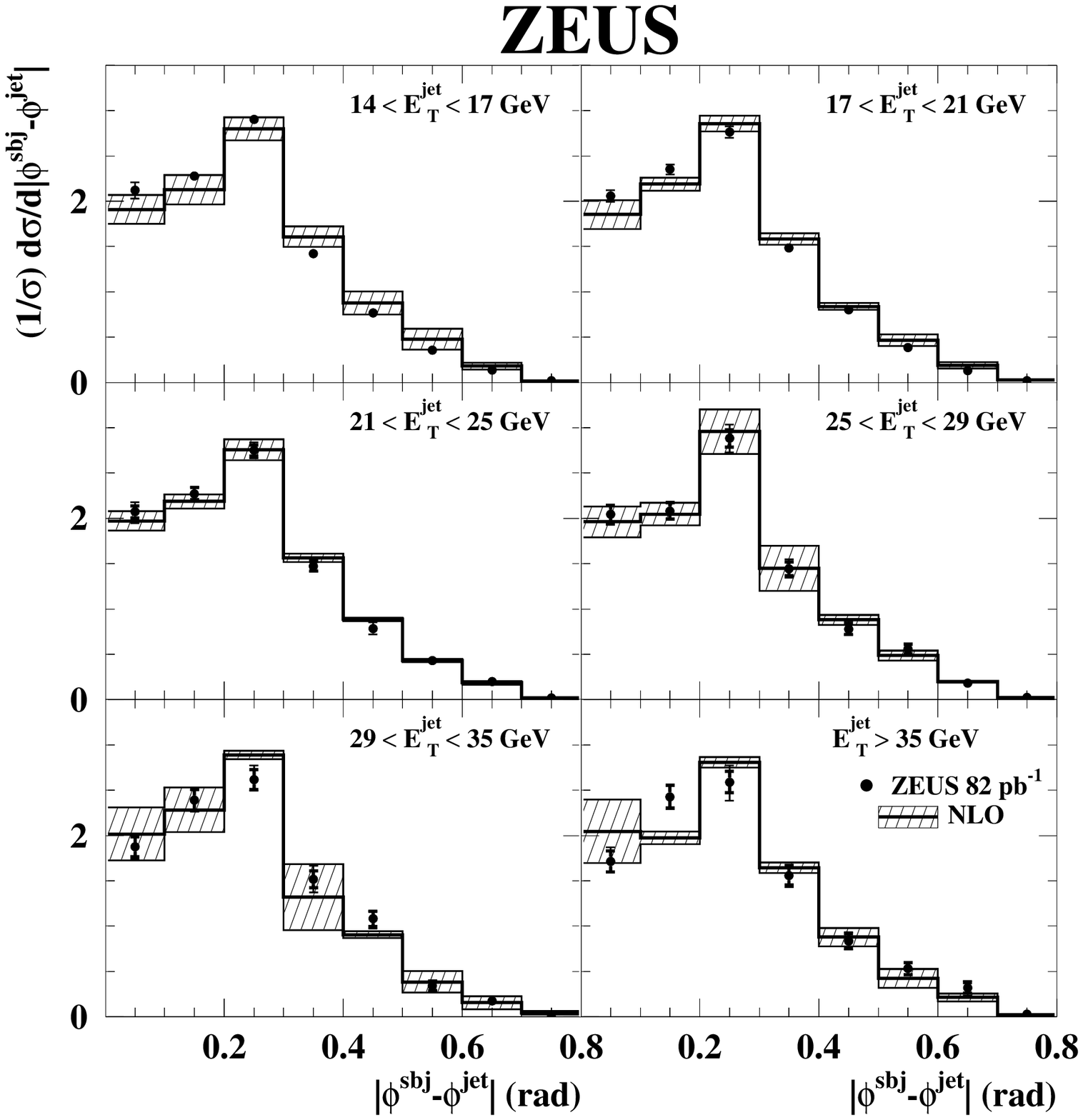,width=18cm}}}
\end{picture}
\caption
{\it 
Measured normalised differential subjet cross sections (dots) for jets
with $\etjet>14$ GeV and $-1<\etajet<2.5$ which have two subjets for
$\yc=0.05$ in the kinematic region given by $\q2>125$~\gev$^2$
as functions of $|\phisbj-\phijet|$ in different regions of
$\etjet$. 
Other details are as in the caption to Fig.~\ref{fig2}.
}
\label{fig7}
\vfill
\end{figure}

\begin{figure}[p]
\vfill
\setlength{\unitlength}{1.0cm}
\begin{picture} (18.0,15.0)
\put (0.0,0.0){\centerline{\epsfig{figure=\figdir 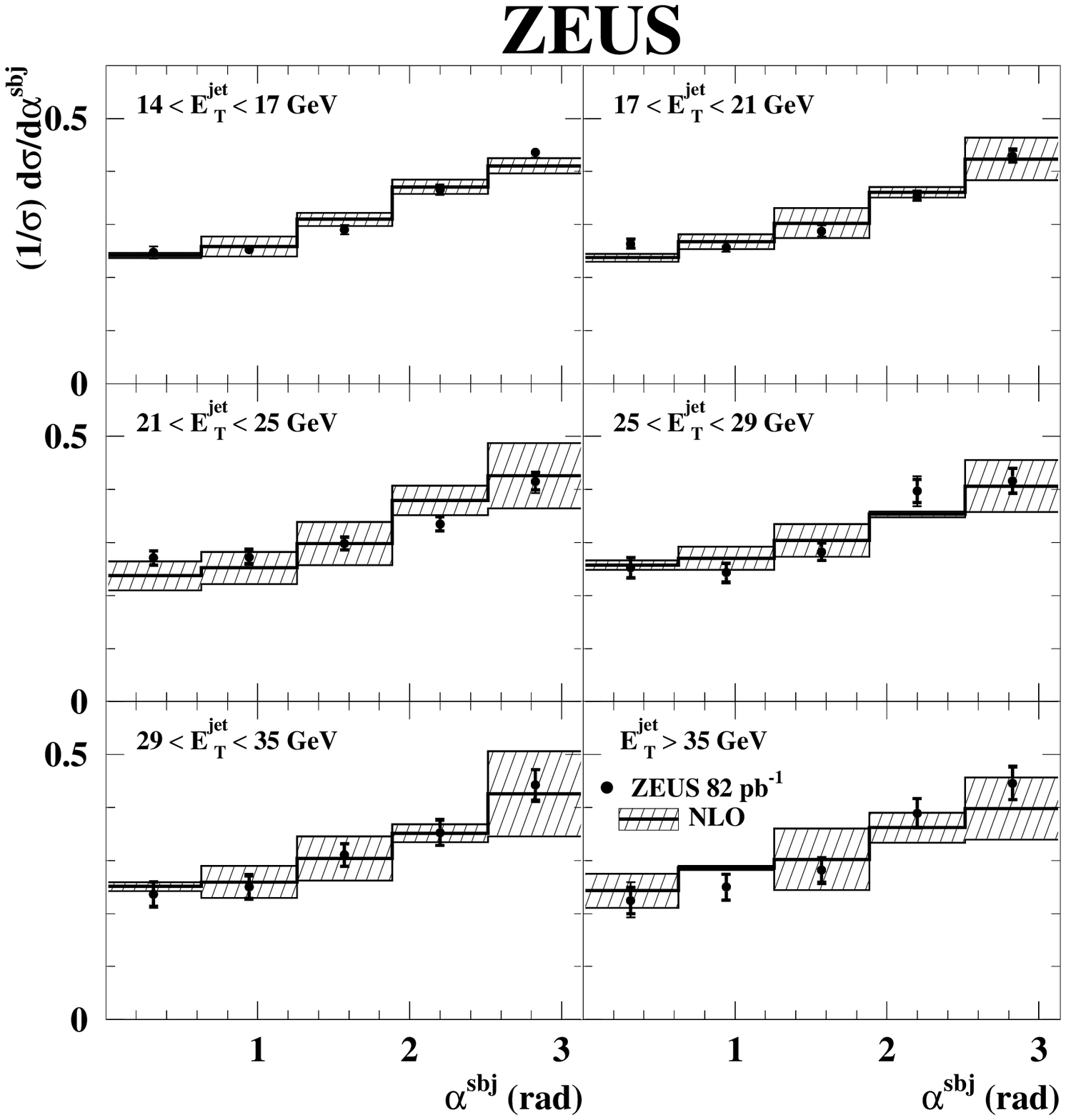,width=18cm}}}
\end{picture}
\caption
{\it 
Measured normalised differential subjet cross sections (dots) for jets
with $\etjet>14$ GeV and $-1<\etajet<2.5$ which have two subjets for
$\yc=0.05$ in the kinematic region given by $\q2>125$~\gev$^2$
as functions of $\asbj$ in different regions of
$\etjet$. 
Other details are as in the caption to Fig.~\ref{fig2}.
}
\label{fig8}
\vfill
\end{figure}

\begin{figure}[p]
\vfill
\setlength{\unitlength}{1.0cm}
\begin{picture} (18.0,15.0)
\put (0.0,0.0){\centerline{\epsfig{figure=\figdir 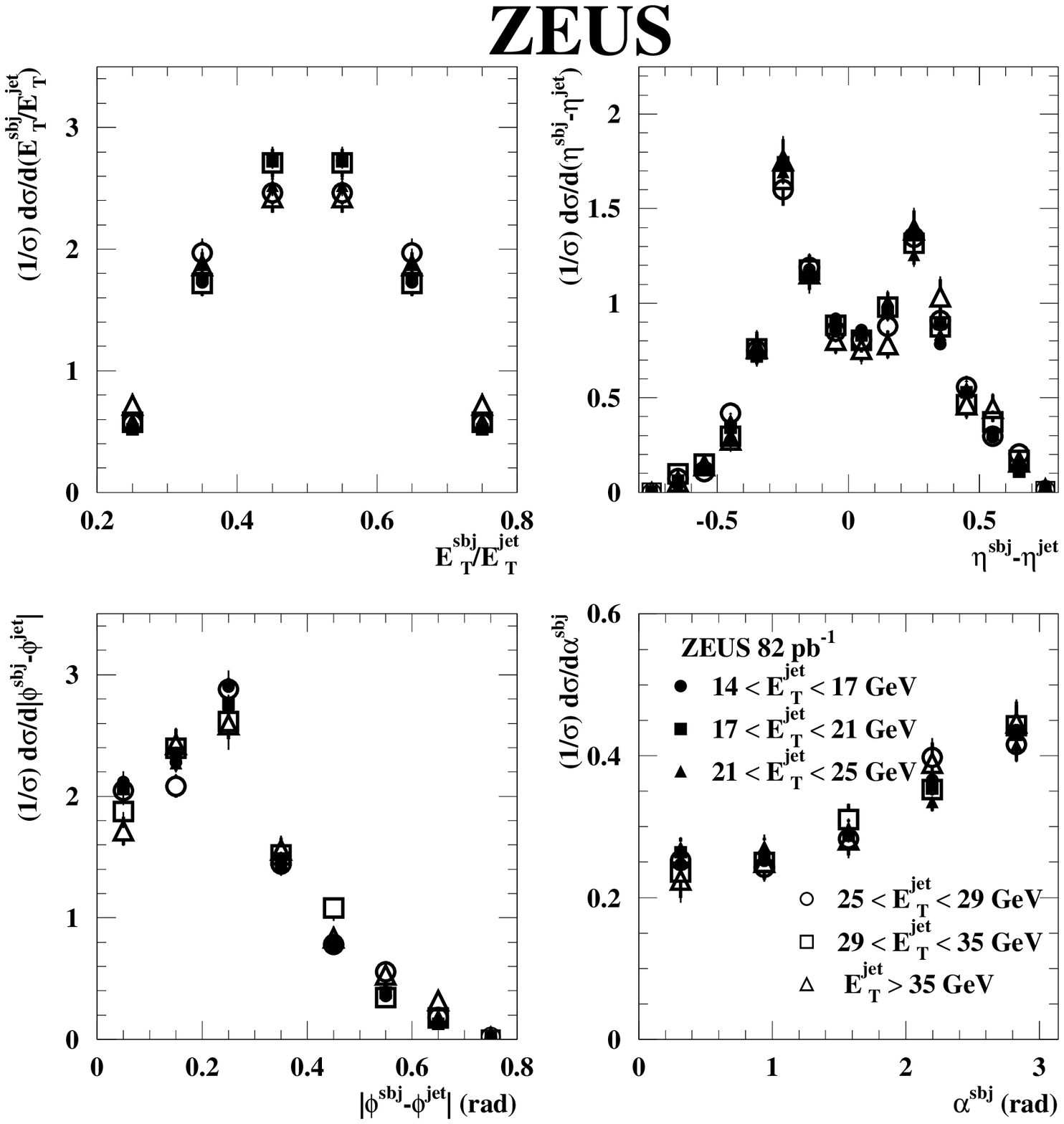,width=20cm}}}
\put (6.0,17.4){\bf\small (a)}
\put (9.6,17.4){\bf\small (b)}
\put (6.0,8.4){\bf\small (c)}
\put (15.0,8.4){\bf\small (d)}
\end{picture}
\caption
{\it 
Measured normalised differential subjet cross sections for jets
with $\etjet>14$ GeV and $-1<\etajet<2.5$ which have two subjets for 
$\yc=0.05$ in the kinematic region given by $\q2>125$~\gev$^2$
as functions of (a) $\etsbj/\etjet$, (b) $\etasbj-\etajet$, (c)
$|\phisbj-\phijet|$ and (d) $\asbj$ in different regions of
$\etjet$. Details concerning the error bars are as in the caption to
Fig.~\ref{fig2}.
}
\label{fig17}
\vfill
\end{figure}

\begin{figure}[p]
\vfill
\setlength{\unitlength}{1.0cm}
\begin{picture} (18.0,15.0)
\put (0.0,0.0){\centerline{\epsfig{figure=\figdir 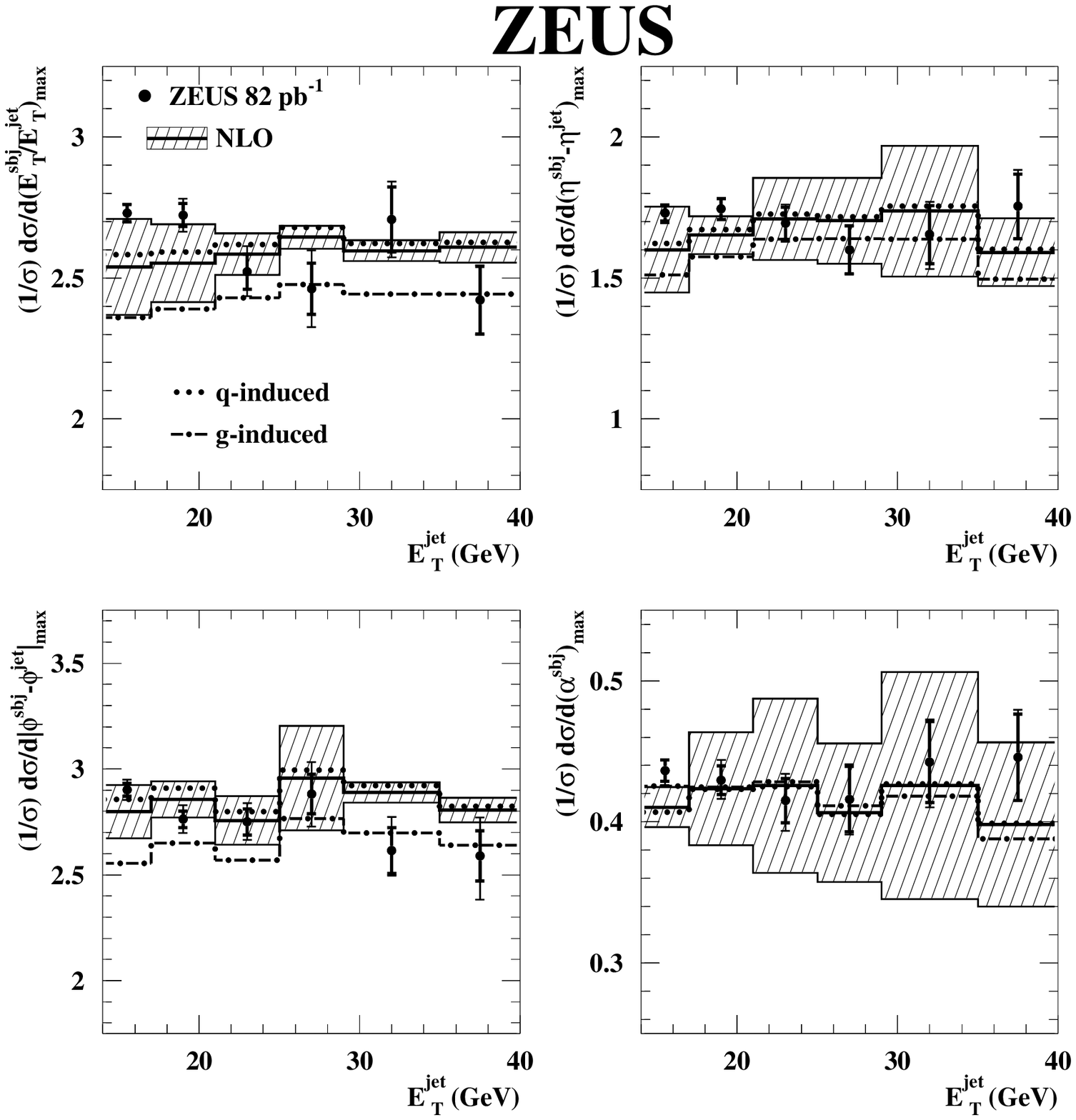,width=20cm}}}
\put (6.0,17.4){\bf\small (a)}
\put (9.6,17.4){\bf\small (b)}
\put (6.0,8.4){\bf\small (c)}
\put (9.6,8.4){\bf\small (d)}
\end{picture}
\caption
{\it 
Maximum of the measured normalised differential (a) $\etsbj/\etjet$, (b)
$\etasbj-\etajet$, (c) $|\phisbj-\phijet|$ and (d) $\asbj$ subjet
cross sections (dots) for jets with $\etjet>14$ GeV and $-1<\etajet<2.5$
which have two subjets for $\yc=0.05$ in the kinematic region given by
$\q2>125$~\gev$^2$ as a function of $\etjet$. For comparison, the NLO
predictions for quark- (dotted histograms) and gluon-induced (dot-dashed
histograms) processes are also shown separately.
Other details are as in the caption to Fig.~\ref{fig2}.
}
\label{fig20}
\vfill
\end{figure}

Figures~\ref{fig5} to \ref{fig8} show the normalised differential
subjet cross sections in different regions of $\etjet$. Even though
the mean subjet multiplicity decreases with increasing
$\etjet$~\cite{pl:b558:41}, the measured normalised differential
subjet cross sections have very similar shapes in all $\etjet$ regions
for all the observables considered. This means that the subjet
topology does not change significantly with $\etjet$. This is better
illustrated in Fig.~\ref{fig17}, where the data for all $\etjet$
regions are plotted together. In particular, it is observed that
the maximum of each measured normalised cross section in every region
of $\etjet$ occurs in the same bin of the distribution. To
quantify the $\etjet$ dependence more precisely, Fig.~\ref{fig20}
shows the maximum value of the measured normalised cross section for
each observable as a function of $\etjet$ together with the NLO
predictions. The spread of the measured maximum values of the
normalised cross sections is $\pm (4-6)\%$. For each observable, 
the scaling behaviour of the normalised differential subjet cross
sections is clearly observed and in agreement with the expectation
that the splitting functions depend weakly on the energy scale. The
NLO QCD calculations are in agreement with the data and support this
observation. 

\begin{figure}[p]
\vfill
\setlength{\unitlength}{1.0cm}
\begin{picture} (18.0,15.0)
\put (0.0,0.0){\centerline{\epsfig{figure=\figdir 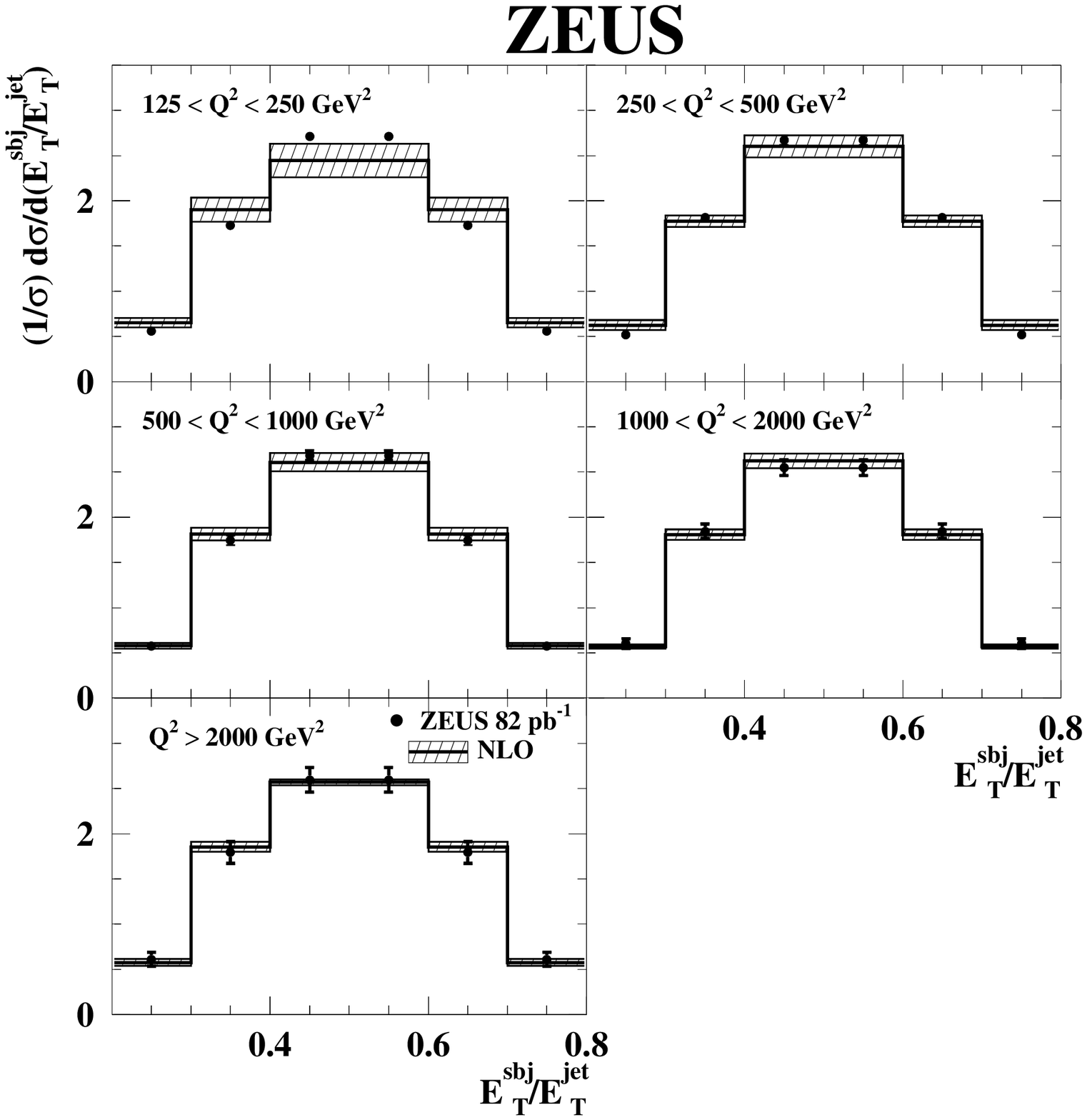,width=18cm}}}
\end{picture}
\caption
{\it 
Measured normalised differential subjet cross sections (dots) for jets
with $\etjet>14$ GeV and $-1<\etajet<2.5$ which have two subjets for
$\yc=0.05$ in the kinematic region given by $\q2>125$~\gev$^2$
as functions of $\etsbj/\etjet$ in different regions of
$\q2$. 
Other details are as in the caption to Fig.~\ref{fig2}.
}
\label{fig9}
\vfill
\end{figure}

\begin{figure}[p]
\vfill
\setlength{\unitlength}{1.0cm}
\begin{picture} (18.0,15.0)
\put (0.0,0.0){\centerline{\epsfig{figure=\figdir 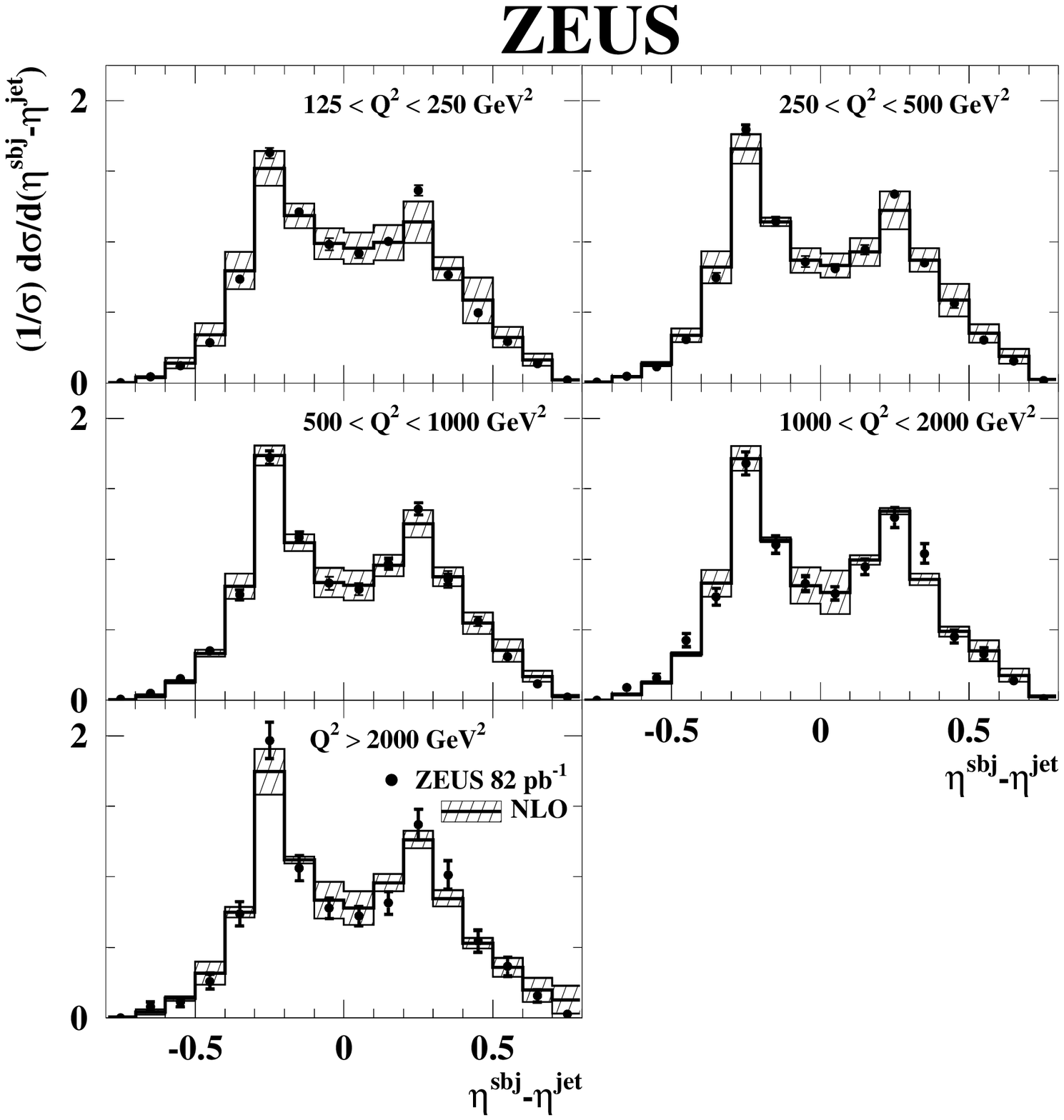,width=18cm}}}
\end{picture}
\caption
{\it 
Measured normalised differential subjet cross sections (dots) for jets
with $\etjet>14$ GeV and $-1<\etajet<2.5$ which have two subjets for
$\yc=0.05$ in the kinematic region given by $\q2>125$~\gev$^2$
as functions of $\etasbj-\etajet$ in different regions of
$\q2$.
Other details are as in the caption to Fig.~\ref{fig2}.
}
\label{fig10}
\vfill
\end{figure}

\begin{figure}[p]
\vfill
\setlength{\unitlength}{1.0cm}
\begin{picture} (18.0,15.0)
\put (0.0,0.0){\centerline{\epsfig{figure=\figdir 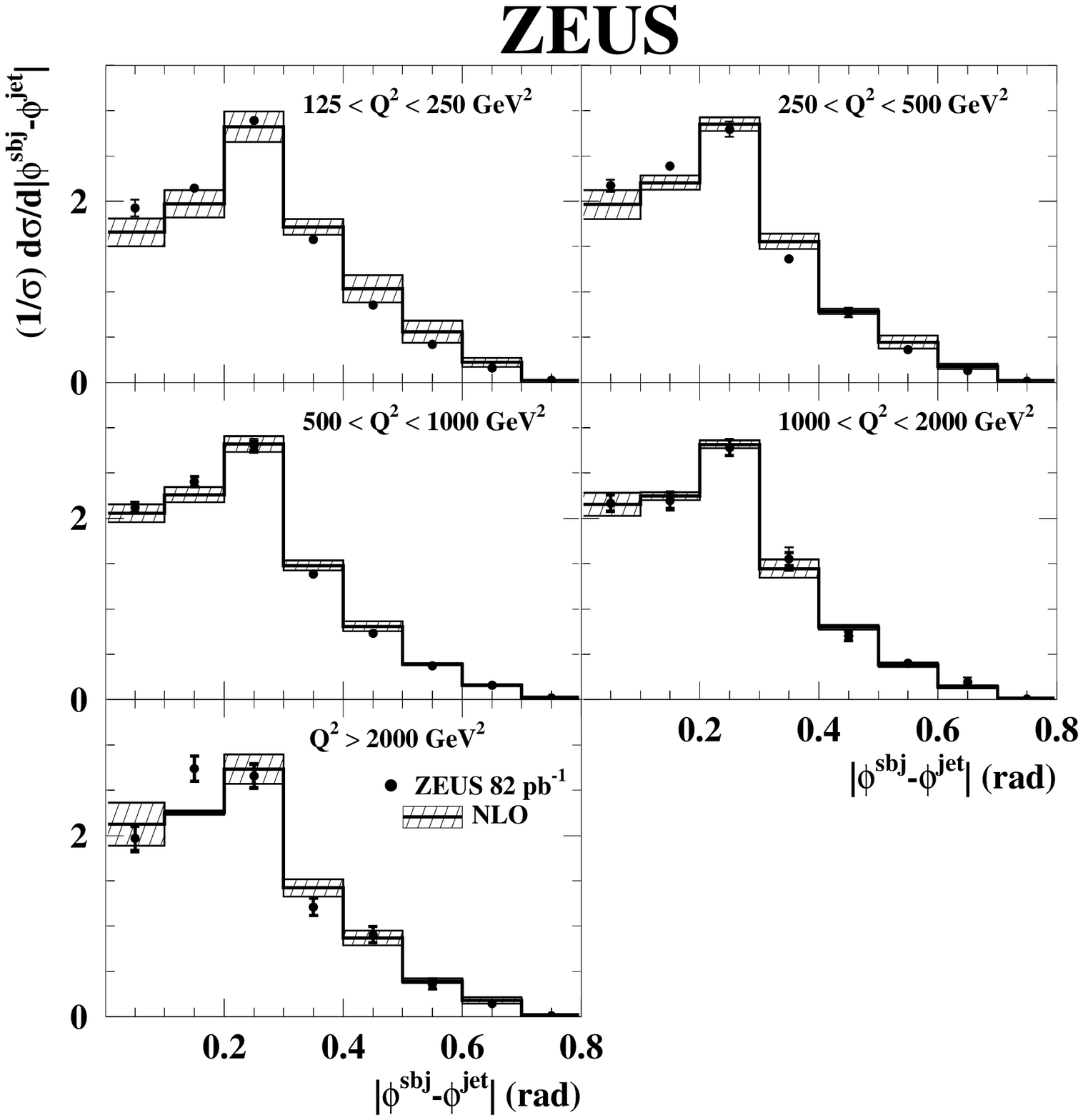,width=18cm}}}
\end{picture}
\caption
{\it 
Measured normalised differential subjet cross sections (dots) for jets
with $\etjet>14$ GeV and $-1<\etajet<2.5$ which have two subjets for
$\yc=0.05$ in the kinematic region given by $\q2>125$~\gev$^2$
as functions of $|\phisbj-\phijet|$ in different regions of
$\q2$.
Other details are as in the caption to Fig.~\ref{fig2}.
}
\label{fig11}
\vfill
\end{figure}

\begin{figure}[p]
\vfill
\setlength{\unitlength}{1.0cm}
\begin{picture} (18.0,15.0)
\put (0.0,0.0){\centerline{\epsfig{figure=\figdir 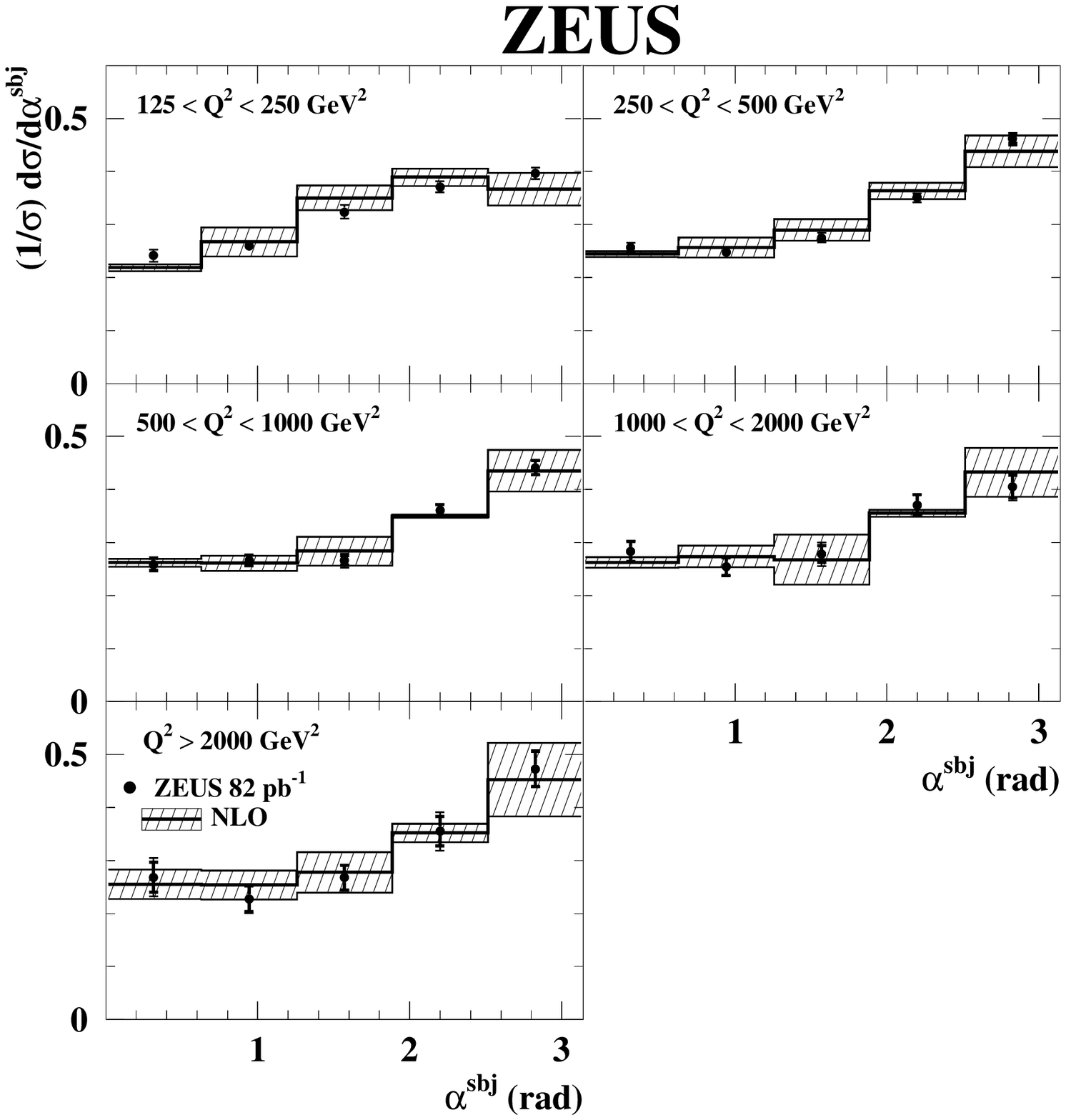,width=18cm}}}
\end{picture}
\caption
{\it 
Measured normalised differential subjet cross sections (dots) for jets
with $\etjet>14$ GeV and $-1<\etajet<2.5$ which have two subjets for
$\yc=0.05$ in the kinematic region given by $\q2>125$~\gev$^2$
as functions of $\asbj$ in different regions of
$\q2$.
Other details are as in the caption to Fig.~\ref{fig2}.
}
\label{fig12}
\vfill
\end{figure}

\begin{figure}[p]
\vfill
\setlength{\unitlength}{1.0cm}
\begin{picture} (18.0,15.0)
\put (0.0,0.0){\centerline{\epsfig{figure=\figdir 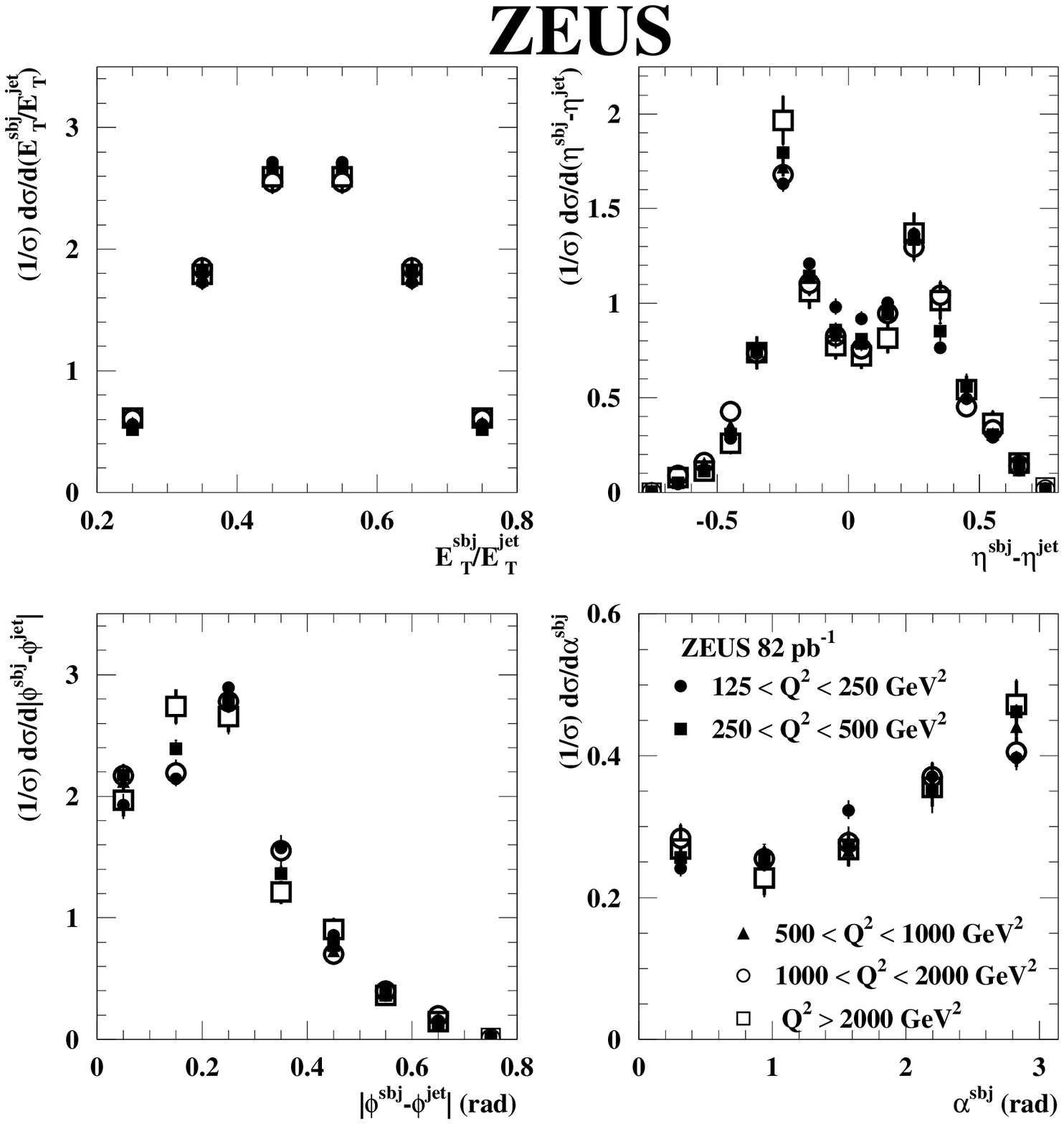,width=20cm}}}
\put (6.0,17.4){\bf\small (a)}
\put (9.6,17.4){\bf\small (b)}
\put (6.0,8.4){\bf\small (c)}
\put (15.0,8.4){\bf\small (d)}
\end{picture}
\caption
{\it 
Measured normalised differential subjet cross sections for jets
with $\etjet>14$ GeV and $-1<\etajet<2.5$ which have two subjets for 
$\yc=0.05$ in the kinematic region given by $\q2>125$~\gev$^2$
as functions of (a) $\etsbj/\etjet$, (b) $\etasbj-\etajet$, (c)
$|\phisbj-\phijet|$ and (d) $\asbj$ in different regions of
$\q2$.
Details concerning the error bars are as in the caption to
Fig.~\ref{fig2}.
}
\label{fig18}
\vfill
\end{figure}

\begin{figure}[p]
\vfill
\setlength{\unitlength}{1.0cm}
\begin{picture} (18.0,15.0)
\put (0.0,0.0){\centerline{\epsfig{figure=\figdir 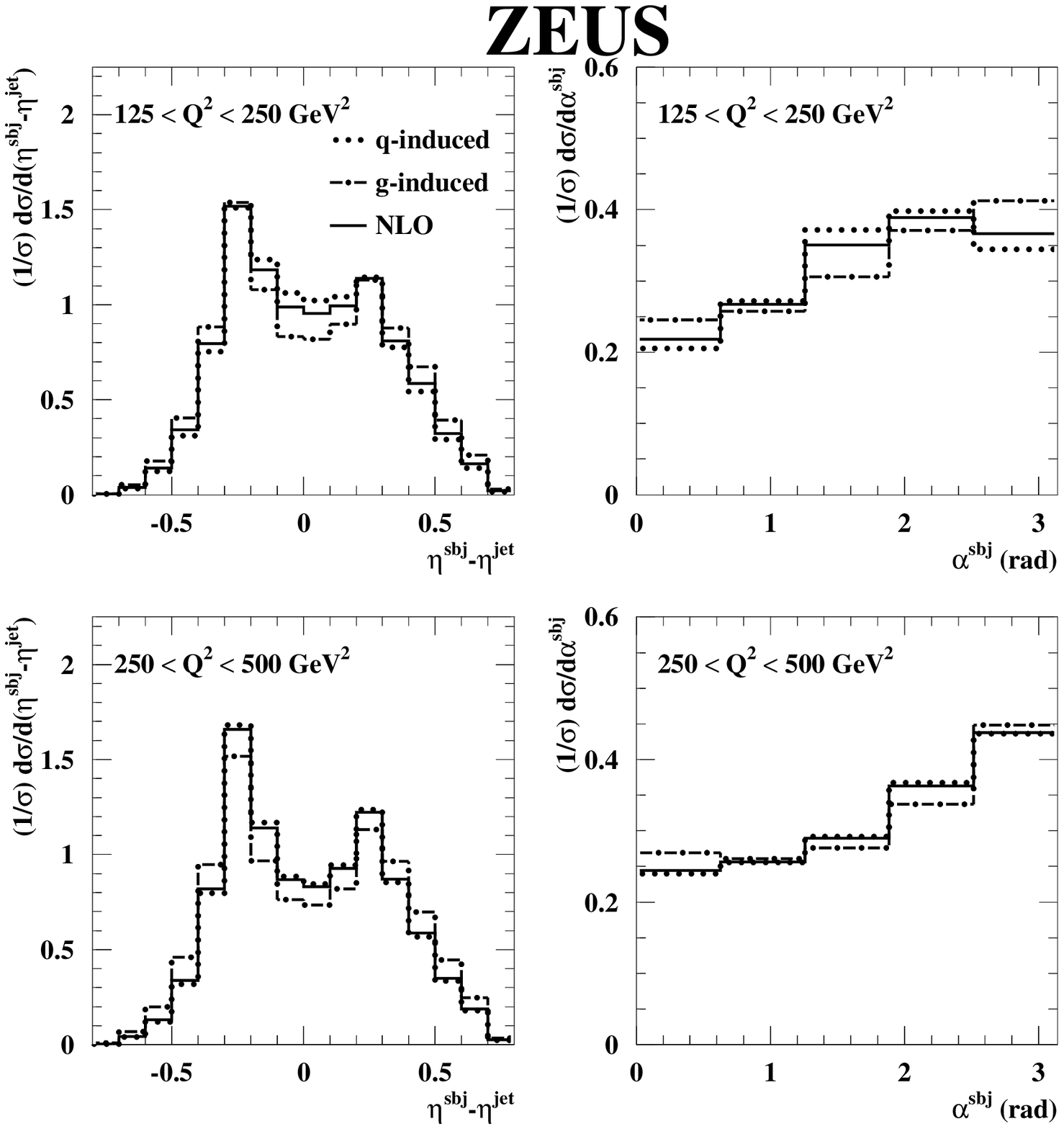,width=20cm}}}
\put (6.0,17.4){\bf\small (a)}
\put (15.0,17.4){\bf\small (b)}
\put (6.0,8.4){\bf\small (c)}
\put (15.0,8.4){\bf\small (d)}
\end{picture}
\caption
{\it 
Predicted normalised differential subjet cross sections (solid
histograms) for jets
with $\etjet>14$ GeV and $-1<\etajet<2.5$ which have two subjets for 
$\yc=0.05$ in the kinematic region given by $\q2>125$~\gev$^2$
as functions of (a,c) $\etasbj-\etajet$ and (b,d) $\asbj$ in
different regions of $\q2$. The NLO predictions for
quark- (dotted histograms) and gluon-induced (dot-dashed histograms)
processes are also shown separately.
}
\label{fig23}
\vfill
\end{figure}

\begin{figure}[p]
\vfill
\setlength{\unitlength}{1.0cm}
\begin{picture} (18.0,15.0)
\put (0.0,0.0){\centerline{\epsfig{figure=\figdir 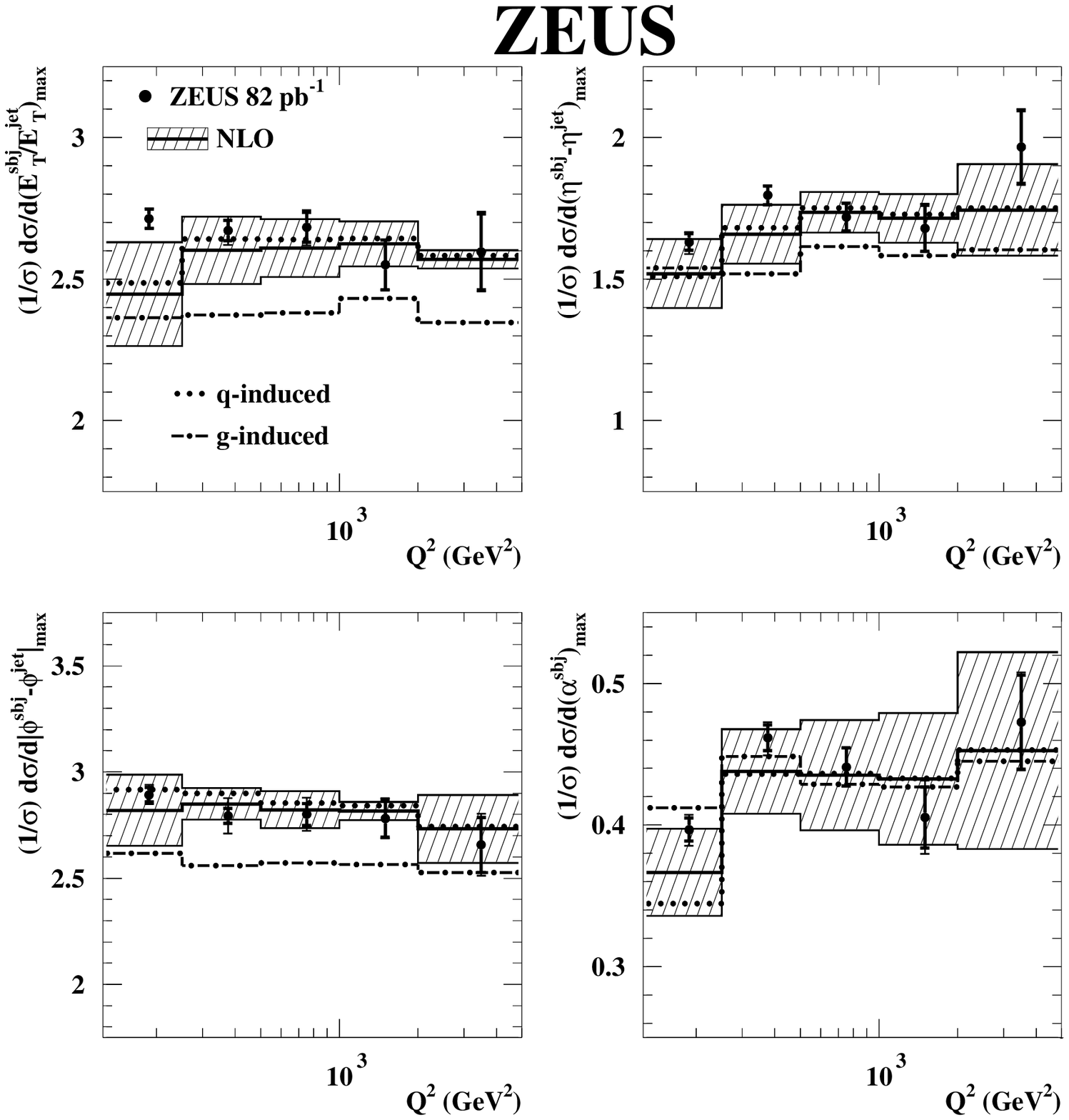,width=20cm}}}
\put (6.0,17.4){\bf\small (a)}
\put (9.6,17.4){\bf\small (b)}
\put (6.0,8.4){\bf\small (c)}
\put (9.6,8.4){\bf\small (d)}
\end{picture}
\caption
{\it 
Maximum of the measured normalised differential (a) $\etsbj/\etjet$, (b)
$\etasbj-\etajet$, (c) $|\phisbj-\phijet|$ and (d) $\asbj$ subjet
cross sections (dots) for jets with $\etjet>14$ GeV and $-1<\etajet<2.5$
which have two subjets for $\yc=0.05$ in the kinematic region given by
$\q2>125$~\gev$^2$ as a function of $\q2$. For comparison, the NLO
predictions for quark- (dotted histograms) and gluon-induced (dot-dashed
histograms) processes are also shown separately.
Other details are as in the caption to Fig.~\ref{fig2}.
}
\label{fig21}
\vfill
\end{figure}

Figures~\ref{fig9} to \ref{fig12} show the normalised differential
subjet cross sections in different regions of $\q2$. In this case, it
is observed that while the shape of the $\etsbj/\etjet$ distribution
does not change significantly with $\q2$, some dependence can be seen
in the other observables. For example, the dip in the $\etasbj-\etajet$
distribution is shallower for $125<\q2<250$~\g2\ than at higher $\q2$
and the shape of the $\asbj$ distribution for $125<\q2<250$~\g2\ is
somewhat different than for the other regions (see
Fig.~\ref{fig18}). These features of the data are reasonably reproduced
by the NLO QCD calculations and understood as a combination of two
effects: the fraction of gluon-induced events is predicted to be
$32\%$ for $125<\q2<250$~\g2\ and below $14\%$ for higher $\q2$; the
shape of the normalised cross sections as functions of
$\etasbj-\etajet$ and $\asbj$ changes from the region
$125<\q2<250$~\g2\ to $250<\q2<500$~\g2\ (see Fig.~\ref{fig23}) 
for quark- and gluon-induced events. It is observed that the maximum
of each measured normalised cross section in every region of $\q2$
occurs in the same bin of the distribution, except for
$|\phisbj-\phijet|$ in the highest-$\q2$ region. Figure~\ref{fig21}
shows the maximum\footnote{For the $|\phisbj-\phijet|$ distribution,
  the same bin has been used for consistency.} value of the measured
normalised cross section for each observable as a function of $\q2$
together with the NLO predictions. The spread of the measured maximum
values of the normalised cross sections as functions of
$\etsbj/\etjet$ and $|\phisbj-\phijet|$ is $\pm (3-4)\%$. On the other
hand, the measured and predicted maximum values for the normalised
cross sections as functions of $\etasbj-\etajet$ and $\asbj$ exhibit a
step-like behaviour between the lowest-$\q2$ region and the rest. 

\begin{figure}[p]
\vfill
\setlength{\unitlength}{1.0cm}
\begin{picture} (18.0,15.0)
\put (0.0,0.0){\centerline{\epsfig{figure=\figdir 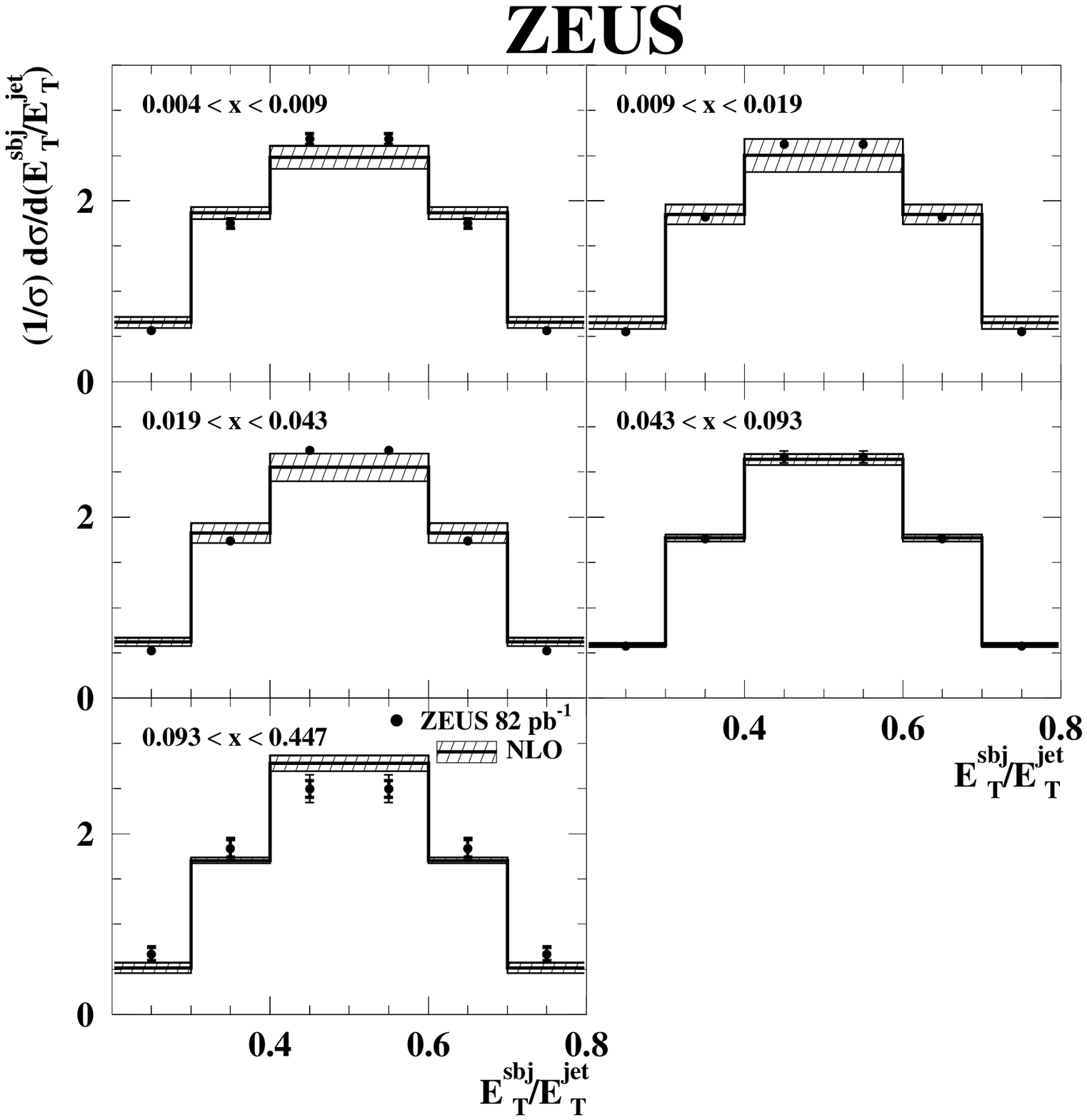,width=18cm}}}
\end{picture}
\caption
{\it 
Measured normalised differential subjet cross sections (dots) for jets
with $\etjet>14$ GeV and $-1<\etajet<2.5$ which have two subjets for
$\yc=0.05$ in the kinematic region given by $\q2>125$~\gev$^2$
as functions of $\etsbj/\etjet$ in different regions of
$x$.
Other details are as in the caption to Fig.~\ref{fig2}.
}
\label{fig13}
\vfill
\end{figure}

\begin{figure}[p]
\vfill
\setlength{\unitlength}{1.0cm}
\begin{picture} (18.0,15.0)
\put (0.0,0.0){\centerline{\epsfig{figure=\figdir 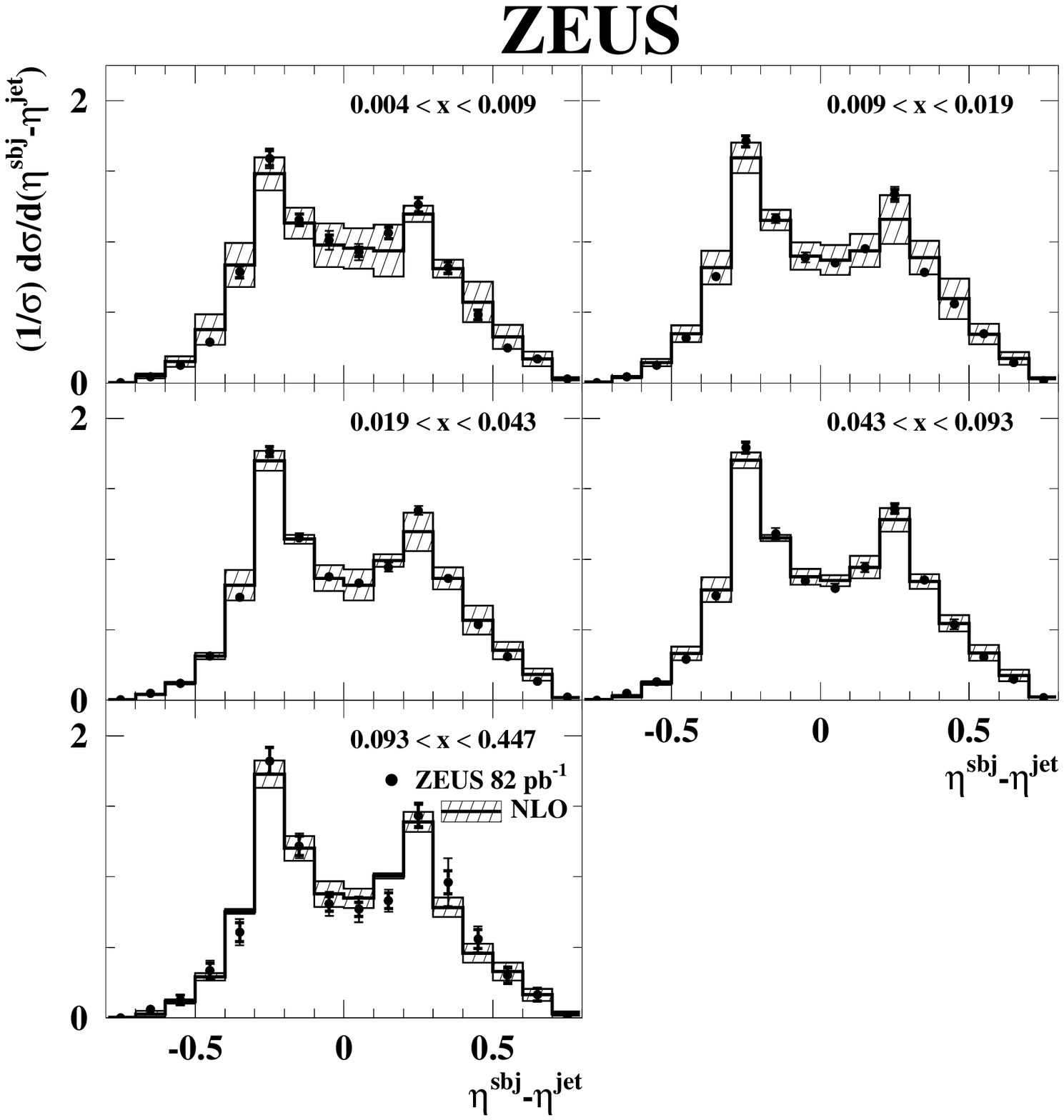,width=18cm}}}
\end{picture}
\caption
{\it 
Measured normalised differential subjet cross sections (dots) for jets
with $\etjet>14$ GeV and $-1<\etajet<2.5$ which have two subjets for
$\yc=0.05$ in the kinematic region given by $\q2>125$~\gev$^2$
as functions of $\etasbj-\etajet$ in different regions of
$x$.
Other details are as in the caption to Fig.~\ref{fig2}.
}
\label{fig14}
\vfill
\end{figure}

\begin{figure}[p]
\vfill
\setlength{\unitlength}{1.0cm}
\begin{picture} (18.0,15.0)
\put (0.0,0.0){\centerline{\epsfig{figure=\figdir 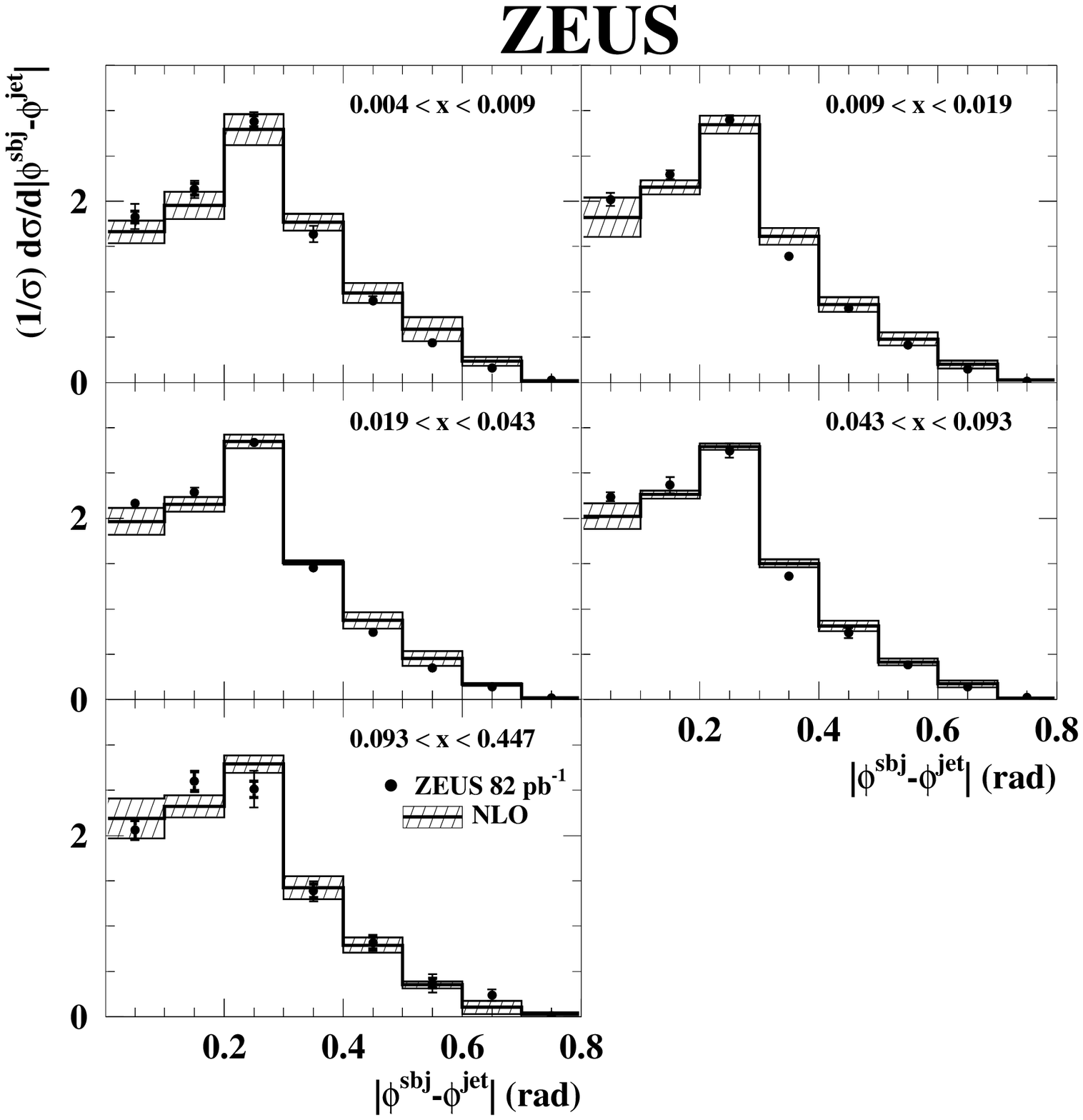,width=18cm}}}
\end{picture}
\caption
{\it 
Measured normalised differential subjet cross sections (dots) for jets
with $\etjet>14$ GeV and $-1<\etajet<2.5$ which have two subjets for
$\yc=0.05$ in the kinematic region given by $\q2>125$~\gev$^2$
as functions of $|\phisbj-\phijet|$ in different regions of
$x$.
Other details are as in the caption to Fig.~\ref{fig2}.
}
\label{fig15}
\vfill
\end{figure}

\begin{figure}[p]
\vfill
\setlength{\unitlength}{1.0cm}
\begin{picture} (18.0,15.0)
\put (0.0,0.0){\centerline{\epsfig{figure=\figdir 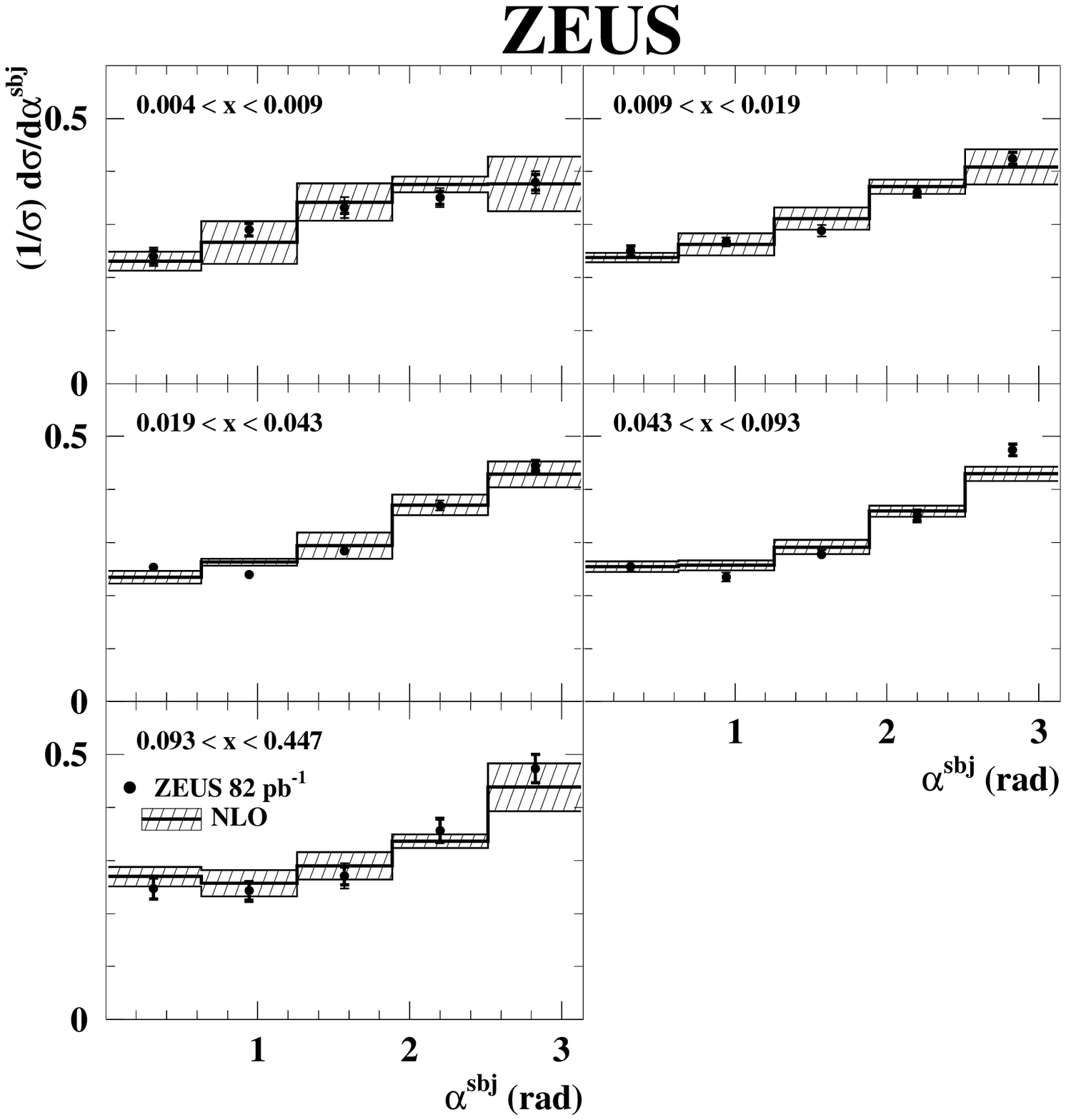,width=18cm}}}
\end{picture}
\caption
{\it 
Measured normalised differential subjet cross sections (dots) for jets
with $\etjet>14$ GeV and $-1<\etajet<2.5$ which have two subjets for
$\yc=0.05$ in the kinematic region given by $\q2>125$~\gev$^2$
as functions of $\asbj$ in different regions of
$x$.
Other details are as in the caption to Fig.~\ref{fig2}.
}
\label{fig16}
\vfill
\end{figure}

\begin{figure}[p]
\vfill
\setlength{\unitlength}{1.0cm}
\begin{picture} (18.0,15.0)
\put (0.0,0.0){\centerline{\epsfig{figure=\figdir 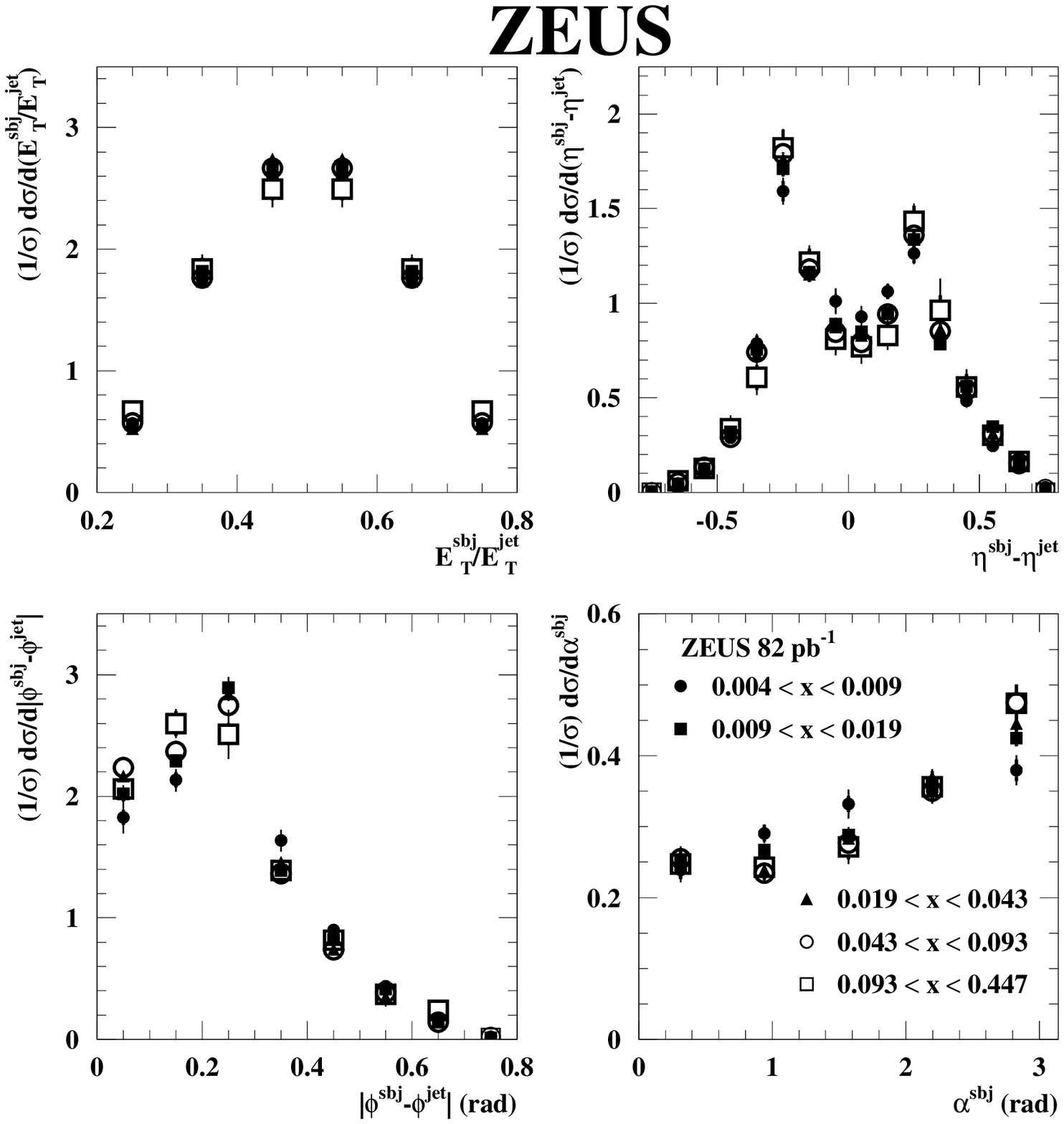,width=20cm}}}
\put (6.0,17.4){\bf\small (a)}
\put (9.6,17.4){\bf\small (b)}
\put (6.0,8.4){\bf\small (c)}
\put (15.0,8.4){\bf\small (d)}
\end{picture}
\caption
{\it 
Measured normalised differential subjet cross sections for jets
with $\etjet>14$ GeV and $-1<\etajet<2.5$ which have two subjets for 
$\yc=0.05$ in the kinematic region given by $\q2>125$~\gev$^2$
as functions of (a) $\etsbj/\etjet$, (b) $\etasbj-\etajet$, (c)
$|\phisbj-\phijet|$ and (d) $\asbj$ in different regions of
$x$.
Details concerning the error bars are as in the caption to
Fig.~\ref{fig2}. 
}
\label{fig19}
\vfill
\end{figure}

\begin{figure}[p]
\vfill
\setlength{\unitlength}{1.0cm}
\begin{picture} (18.0,15.0)
\put (0.0,0.0){\centerline{\epsfig{figure=\figdir 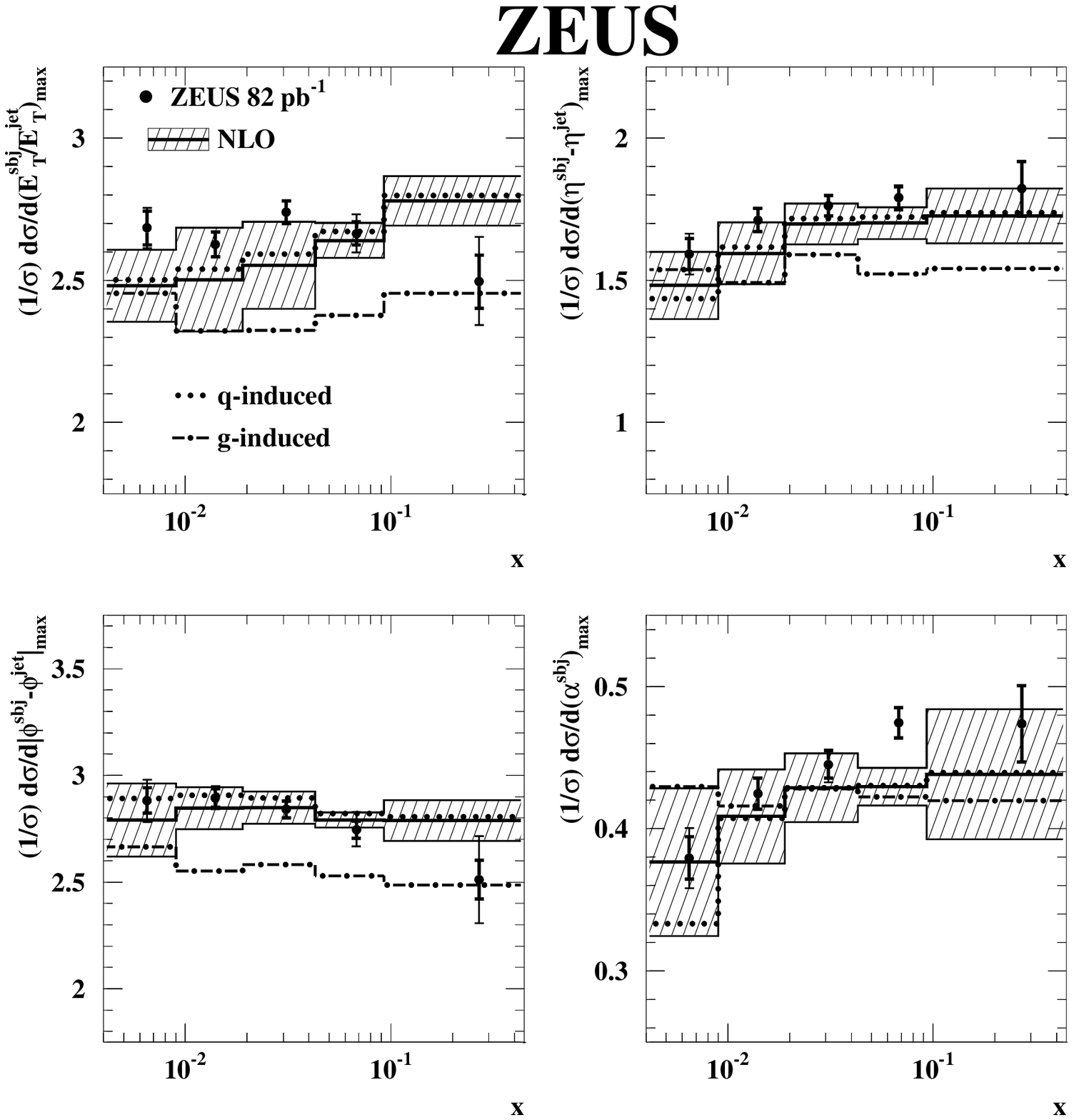,width=20cm}}}
\put (6.0,17.4){\bf\small (a)}
\put (9.6,17.4){\bf\small (b)}
\put (6.0,8.4){\bf\small (c)}
\put (9.6,8.4){\bf\small (d)}
\end{picture}
\caption
{\it 
Maximum of the measured normalised differential (a) $\etsbj/\etjet$, (b)
$\etasbj-\etajet$, (c) $|\phisbj-\phijet|$ and (d) $\asbj$ subjet
cross sections (dots) for jets with $\etjet>14$ GeV and $-1<\etajet<2.5$
which have two subjets for $\yc=0.05$ in the kinematic region given by
$\q2>125$~\gev$^2$ as a function of $x$. For comparison, the NLO
predictions for quark- (dotted histograms) and gluon-induced (dot-dashed
histograms) processes are also shown separately.
Other details are as in the caption to Fig.~\ref{fig2}.
}
\label{fig22}
\vfill
\end{figure}

\begin{figure}[p]
\vfill
\setlength{\unitlength}{1.0cm}
\begin{picture} (18.0,15.0)
\put (0.0,0.0){\centerline{\epsfig{figure=\figdir 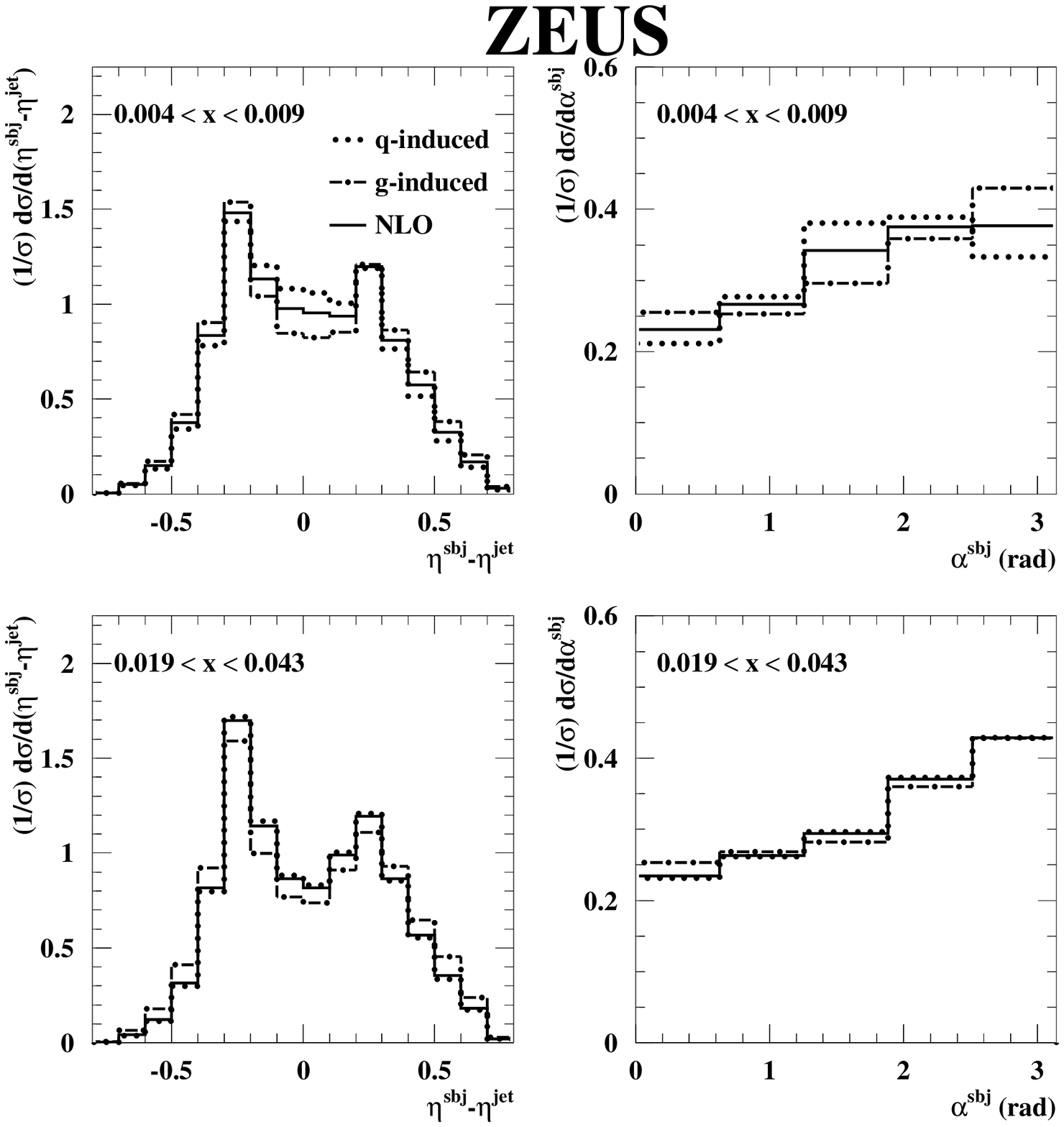,width=20cm}}}
\put (6.0,17.4){\bf\small (a)}
\put (15.0,17.4){\bf\small (b)}
\put (6.0,8.4){\bf\small (c)}
\put (15.0,8.4){\bf\small (d)}
\end{picture}
\caption
{\it 
Predicted normalised differential subjet cross sections (solid
histograms) for jets
with $\etjet>14$ GeV and $-1<\etajet<2.5$ which have two subjets for 
$\yc=0.05$ in the kinematic region given by $\q2>125$~\gev$^2$
as functions of (a,c) $\etasbj-\etajet$ and (b,d) $\asbj$ in
different regions of $x$. The NLO predictions for
quark- (dotted histograms) and gluon-induced (dot-dashed histograms)
processes are also shown separately.
}
\label{fig24}
\vfill
\end{figure}

Figures~\ref{fig13} to \ref{fig16} show the normalised differential
subjet cross sections in different regions of $x$. Figure~\ref{fig19}
shows the data for all $x$ regions plotted together. It is observed
that the maximum of each measured normalised cross section in every
region of $x$ occurs in the same bin of the distribution, except for
$|\phisbj-\phijet|$ in the highest $x$ region. Figure~\ref{fig22}
shows the maximum$^3$ value of the measured
normalised cross section for each observable as a function of $x$. 
The shape of the $\etsbj/\etjet$ measured distribution does not change
significantly with $x$, whereas some dependence is expected (see
Fig.~\ref{fig22}a). The dependence of the $\etasbj-\etajet$ and
$\asbj$ distributions with $x$ exhibits features similar to those
observed in the study of the $\q2$ dependence; in particular, the
maximum values (see Figs.~\ref{fig22}b and \ref{fig22}d) exhibit a
monotonic increase as $x$ increases, which is reasonably reproduced by
the calculations. As discussed previously, these features are 
understood as a combination of two effects: a decrease of the
predicted fraction of gluon-induced events from $44\%$ for
$0.004<x<0.009$ to $6\%$ for $x>0.093$ and the change in shape of the
normalised cross sections for quark- and gluon-induced processes
as $x$ increases (see Fig.~\ref{fig24}).

\begin{figure}[p]
\vfill
\setlength{\unitlength}{1.0cm}
\begin{picture} (18.0,15.0)
\put (0.0,0.0){\centerline{\epsfig{figure=\figdir 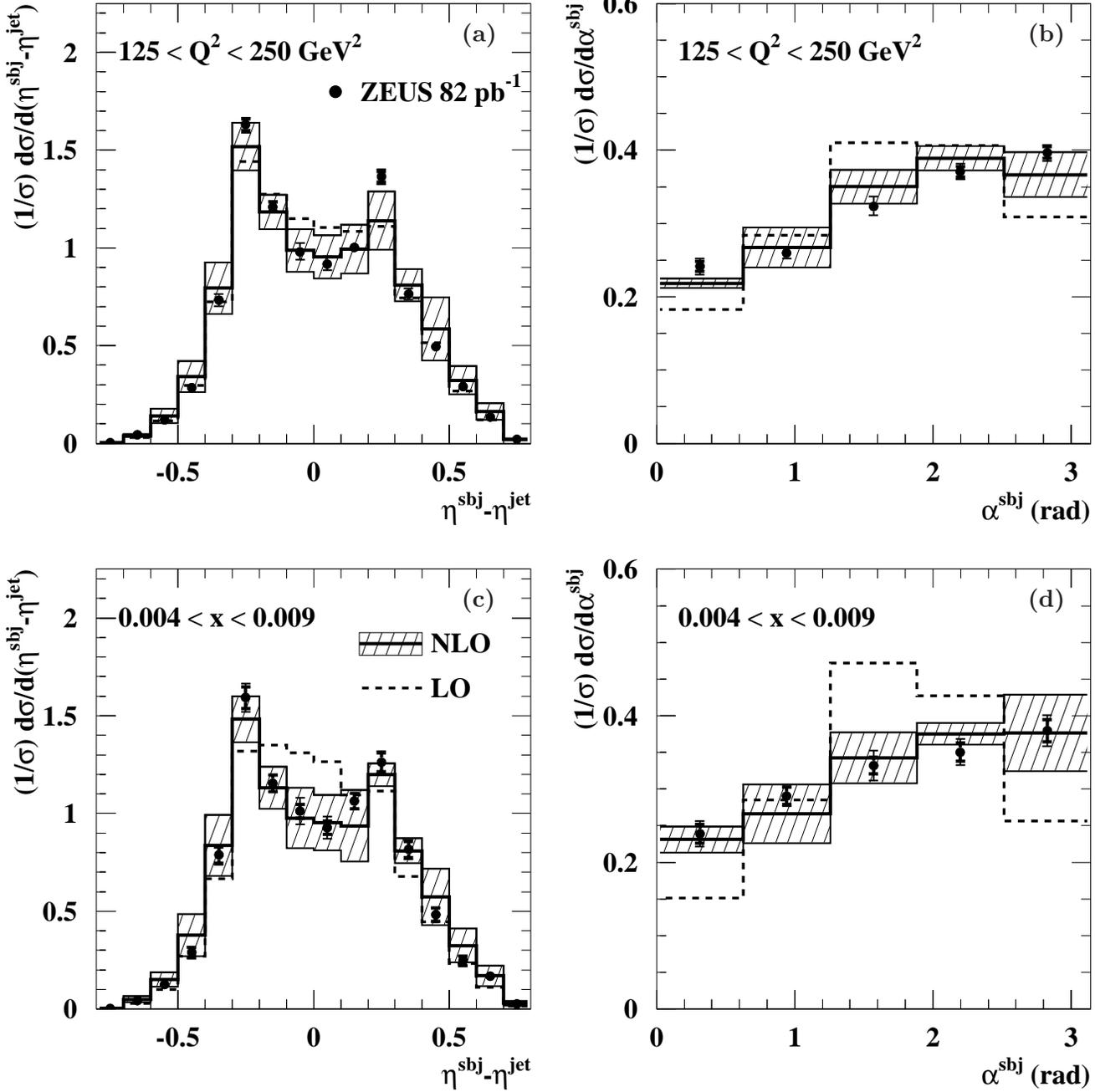,width=20cm}}}
\put (6.0,17.4){\bf\small (a)}
\put (15.0,17.4){\bf\small (b)}
\put (6.0,8.4){\bf\small (c)}
\put (15.0,8.4){\bf\small (d)}
\end{picture}
\caption
{\it 
Measured normalised differential subjet cross sections (dots) for jets
with $\etjet>14$ GeV and $-1<\etajet<2.5$ which have two subjets for 
$\yc=0.05$ in restricted $\q2$ and $x$ regions as functions of (a,c)
$\etasbj-\etajet$ and (b,d) $\asbj$. The NLO (solid histograms)
and LO (dashed histograms) calculations are also shown. The hatched
bands represent the NLO theoretical uncertainty.
}
\label{fig25}
\vfill
\end{figure}

To investigate the origin of the change in shape of the normalised
differential cross sections between the lowest and higher $\q2$ and
$x$ regions, LO and NLO calculations were compared. The
most dramatic change is observed when restricting the kinematic region
to $125<\q2<250$~\g2\ or $0.004<x<0.009$ (see Fig.~\ref{fig25});
the LO calculation of the $\etasbj-\etajet$ distribution does not
exhibit a two-peak structure as seen in the NLO prediction and in the
data. In addition, the LO calculation of the $\asbj$ distribution
peaks at $\asbj\sim \pi/2$ in contrast with the NLO prediction and the
data. This proves that the NLO QCD radiative corrections are
responsible for these variations in shape and necessary for describing
the data.

In summary, while the shapes of the normalised differential cross
sections show only a weak dependence on $\etjet$, their dependence
on $\q2$ and $x$ have some prominent features at low $\q2$ or $x$. The
weak dependence on $\etjet$ is consistent with the expected scaling
behaviour of the splitting functions; however, the restriction to
low $\q2$ or $x$ values demonstrates that the NLO QCD radiative
corrections are important there. The NLO QCD calculations, which
include the two competing processes $eq\rightarrow eqg$ and
$eg\rightarrow e\qq$ and radiative corrections, adequately reproduce
the measurements.

\section{Summary}

Normalised differential subjet cross sections in inclusive-jet NC DIS
were measured in $ep$ collisions using $81.7$~\pb1\ of data collected
with the ZEUS detector at HERA. The cross sections refer to jets
identified in the laboratory frame with the $\kt$ cluster algorithm in
the longitudinally invariant inclusive mode and selected with
$\etjet>14$ GeV and $-1<\etajet<2.5$. The measurements were made for
those jets which have exactly two subjets for $\yc=0.05$ in the
kinematic region defined by $\q2>125$ \g2.

The cross sections were measured as functions of $\etsbj/\etjet$,
$\etasbj-\etajet$, $|\phisbj-\phijet|$ and $\asbj$. The data show that
the two subjets tend to have similar transverse energies and that the
subjet with higher transverse energy tends to be in the rear
direction. This is consistent with the effects of colour coherence
between the initial and final states, which predict that soft parton
radiation is emitted predominantly towards the proton beam direction.

An adequate description of the data is given by NLO QCD
calculations. This means that the pattern of parton radiation as
predicted by QCD reproduces the subjet topology in the
data. Furthermore, the subjet distributions in the data are better
described by the calculations for jets arising from a quark-gluon
pair.

The normalised cross sections show a weak dependence on $\etjet$, in
agreement with the expected scaling behaviour of the splitting
functions. By restricting the measurements to low $\q2$ or $x$
values, significant differences in shape are observed, which can be
primarily attributed to NLO QCD radiative corrections.

\vspace{0.5cm}
\noindent {\Large\bf Acknowledgements}
\vspace{0.3cm}

We thank the DESY Directorate for their strong support and
encouragement. We appreciate the contributions to the construction and
maintenance of the ZEUS detector of many people who are not listed as
authors. The HERA machine group and the DESY computing staff are
especially acknowledged for their success in providing excellent
operation of the collider and the data-analysis environment.

\vfill\eject

\providecommand{\etal}{et al.\xspace}
\providecommand{\coll}{Coll.\xspace}
\catcode`\@=11
\def\@bibitem#1{%
\ifmc@bstsupport
  \mc@iftail{#1}%
    {;\newline\ignorespaces}%
    {\ifmc@first\else.\fi\orig@bibitem{#1}}
  \mc@firstfalse
\else
  \mc@iftail{#1}%
    {\ignorespaces}%
    {\orig@bibitem{#1}}%
\fi}%
\catcode`\@=12

\newpage
\clearpage

\newpage
\clearpage

\newpage
\clearpage

\newpage
\clearpage

\newpage
\clearpage

\newpage
\clearpage

\newpage
\clearpage

\newpage
\clearpage

\newpage
\clearpage

\newpage
\clearpage

\newpage
\clearpage

\newpage
\clearpage

\newpage
\clearpage

\newpage
\clearpage

\newpage
\clearpage

\newpage
\clearpage

\newpage
\clearpage

\newpage
\clearpage

\newpage
\clearpage

\newpage
\clearpage

\newpage
\clearpage

\newpage
\clearpage

\newpage
\clearpage

\newpage
\clearpage

\newpage
\clearpage

\newpage
\clearpage

\end{document}